\documentclass[aps,prl,longbibliography,twocolumn]{revtex4-2}

\usepackage{amsfonts}
\usepackage{amsmath}
\usepackage{graphicx}
\usepackage{overpic}
\usepackage{dsfont}
\usepackage{diagbox}
\usepackage{array}
\usepackage{amssymb}
\usepackage{bbm}
\usepackage{float}
\usepackage{subfigure}
\usepackage{MnSymbol}
\usepackage{url}
\usepackage{tgtermes}  
\usepackage[T1]{fontenc}
\usepackage{manfnt}
\usepackage{booktabs}
\usepackage{xcolor}
\definecolor{darkblue}{HTML}{004D6B}
\definecolor{darkred}{HTML}{8c1515}
\definecolor{darkgreen}{HTML}{006400}
\pagecolor{white}
\usepackage{ulem}
\usepackage{physics}
\usepackage{enumitem}
\usepackage{wasysym}
\usepackage{tikz}
\usetikzlibrary{arrows.meta}
\usepackage{hyperref} 
\hypersetup{
    pdftitle={classic},
	breaklinks,
	colorlinks=true,
	urlcolor=darkred,
	linkcolor=darkblue,
	citecolor=darkgreen
}
\usepackage[noabbrev]{cleveref}
\crefname{figure}{Fig.}{Figs.} 
\Crefname{figure}{Figure}{Figures} 
\crefname{section}{Sec.}{Secs.} 
\Crefname{section}{Section}{Sections}
\crefname{subsection}{Sec.}{Secs.} 
\Crefname{subsection}{Section}{Sections}
\crefname{equation}{Eq.}{Eqs.}
\Crefname{equation}{Equation}{Equations}
\crefname{appendix}{Appendix}{Appendices}
\Crefname{appendix}{Appendix}{Appendices}

\begin{document}

\title{Ergotropy of quantum many-body scars}
\author{Zhaohui Zhi}
\affiliation{The Hong Kong University of Science and Technology (Guangzhou), Nansha, Guangzhou, 511400, Guangdong, China}
\author{Qingyun Qian}
\affiliation{The Hong Kong University of Science and Technology (Guangzhou), Nansha, Guangzhou, 511400, Guangdong, China}
\author{Jin-Guo Liu}
\affiliation{The Hong Kong University of Science and Technology (Guangzhou), Nansha, Guangzhou, 511400, Guangdong, China}
\author{Guo-Yi Zhu}
\email{guoyizhu@hkust-gz.edu.cn}
\affiliation{The Hong Kong University of Science and Technology (Guangzhou), Nansha, Guangzhou, 511400, Guangdong, China}
\date{\today}

\begin{abstract}
Quantum many-body scars break ergodicity and evade thermalization, resulting in sub-volume law scaling entanglement entropy even with high energy density. 
While their quantum correlations and entanglement have been elaborated previously, their capacity in storing extractable energy, quantified by the notion {\it ergotropy}, remains an open question. 
Here we focus on the representative PXP model, and unveil the extensive ergotropy scaling of a family of states interpolating between quantum many-body scars and thermal states, the latter of which are known to be passive with vanishing ergotropy in the thermodynamic limit. A phenomenological relation between ergotropy and entanglement is uncovered, which generalizes the existing free fermion integrable results to an interacting scenario. 
The ergotropy in a dynamical protocol shows that a reset with a global uniform coherent rotation can inject extractable energy, as a proof of principle way to charge a quantum ``battery''. 
Our protocol is tailored for near term Rydberg neutral atom arrays, while also being feasible for other quantum processors.
Our results establish that quantum many-body scars, despite the tiny fraction of the Hilbert space they occupy, can be efficiently exploited for storing extractable energy, and ``scarring'' a many-body system as a promising route for engineering quantum many-body battery.
\end{abstract}

\maketitle


Thermalization is generally expected in isolated quantum many-body systems~\cite{D'Alessio03052016, Lukin_Entanglement_thermalization,Regnault22reviewscar,Huse_MBL_ETH, Popescu2006_thermalization}, where evolved steady states or typical eigenstates become locally indistinguishable from an equilibrated thermal ensemble satisfying all conservation laws, as predicted by the eigenstate thermalization hypothesis (ETH)~\cite{Thermalization_Deutsch_1991,Thermalization_Srednicki_1994,Thermalization_Rigol_2008,QD_Canonical_Tasaki1998, Huse_MBL_ETH}. 
Such thermal states maximize the entropy and are passive: no unitary operation acting on a subsystem can extract energy~\cite{skrzypczyk2015passivity,pusz1978passive, Touil_2022_ergotropy_correlations, Lenard1978_passivity}. Consequently, their ergotropy, defined as the maximal amount of extractable work by performing unitary operations, vanishes in the thermodynamic limit~\cite{francica2020quantum, Touil_2022_ergotropy_correlations}. Therefore, high energy (relative to the ground state) quantum states evading thermalization would be desirable for manipulating energy in a coherent quantum many-body system~\cite{QUantumThermoReviewVinjanampathy2016}, such as engineering quantum batteries~\cite{Campaioli2024, campaioli2017enhancing,ferraro2018high, Alicki2013_ergotropy}, which could possibly be leveraged for quantum information processing~\cite{Kurman2025}. 

Quantum many-body scars (QMBS) represent a remarkable class of dynamical states that circumvent thermalization, first discovered in a quantum quench experiment of Rydberg atom arrays~\cite{Bernien2017} and subsequently identified as a widespread phenomenon in many-body systems~\cite{Papic21reviewscar,Regnault22reviewscar,Moessner23scar,Eta_pairing_Yang1989, XY_scar_Iadecola2019,AKLT_scar_analytical,AKLT_scar_numerical, su2023observation, liang2025observation, zhang2023many, Lin2020_2D_scar, Surace2021_exact_scars}. 
Unlike typical high energy eigenstates, QMBS exhibit sub-volume law entanglement entropy while residing at finite energy density shell~\cite{Papic18reviewScar,Papic21reviewscar,Regnault22reviewscar,AKLT_scar_numerical, AKLT_scar_analytical,XY_scar_Iadecola2019}, suggesting the potential for energy storage and extraction. 
In this way, the quantum quench is turned into a periodic {\it reset} in each cycle of charging energy as an open quantum system method, which should be distinguished from cooling a quantum system~\cite{OBrien21cooling, Wilzeck22cooling, Matthies2024programmable, Koch24cooling}. After reset and relaxation by intrinsic Hamiltonian evolution, the ergotropy of the steady state quantifies the maximal extractable energy.

The ergotropy is intimately connected to entanglement structure~\cite{Yang22coherence,Touil_2022_ergotropy_correlations,Mula2023,Mitra_2025,D'Alessio03052016,MBLquantumbatteries}. Maximally entangled states (such as the random quantum state~\cite{Page_curve}, or thermal state under certain energy constraint\cite{maximal_entropy_principle}) correspond to subsystems at effectively infinite temperature, yielding zero extractable energy. Conversely, low entanglement ground states contain limited energy compared to higher energy shells. This trade-off motivates a thermodynamic study of QMBS: how much energy can be extracted from these exotic, high energy yet low entanglement states?

Here we present an exact diagonalization study of ergotropy in both eigenstates and real time evolution for the PXP model. The QMBS and thermal states are separated from the ensemble of degenerate eigenstates within the same energy shell $E=0$ at the middle of the many-body spectrum, representing ``infinite temperature". We observe a crossover from extensive to sub-extensive ergotropy by tuning the superposition between scar and thermal components, alongside the entanglement structure. 
In the dynamical quantum protocol, we use a coherent rotation to drive the steady state crossing over from QMBS to thermal states, exhibiting qualitatively distinct ergotropy dynamics. 
Furthermore, we show that the relation between unextractable bound energy $Q$ and von Neumann entropy $S_{\rm vN}$: $Q \propto S_{\mathrm{vN}}^2$~\cite{Mula2023,Mitra_2025} can be generalized from free fermion to a strongly {\it interacting} system. This implies that higher entanglement entropy suppresses ergotropy, providing an entanglement based guiding principle to employ QMBS for engineering quantum batteries.

\begin{figure}[tb!]
    \centering
    \includegraphics[width=\columnwidth]{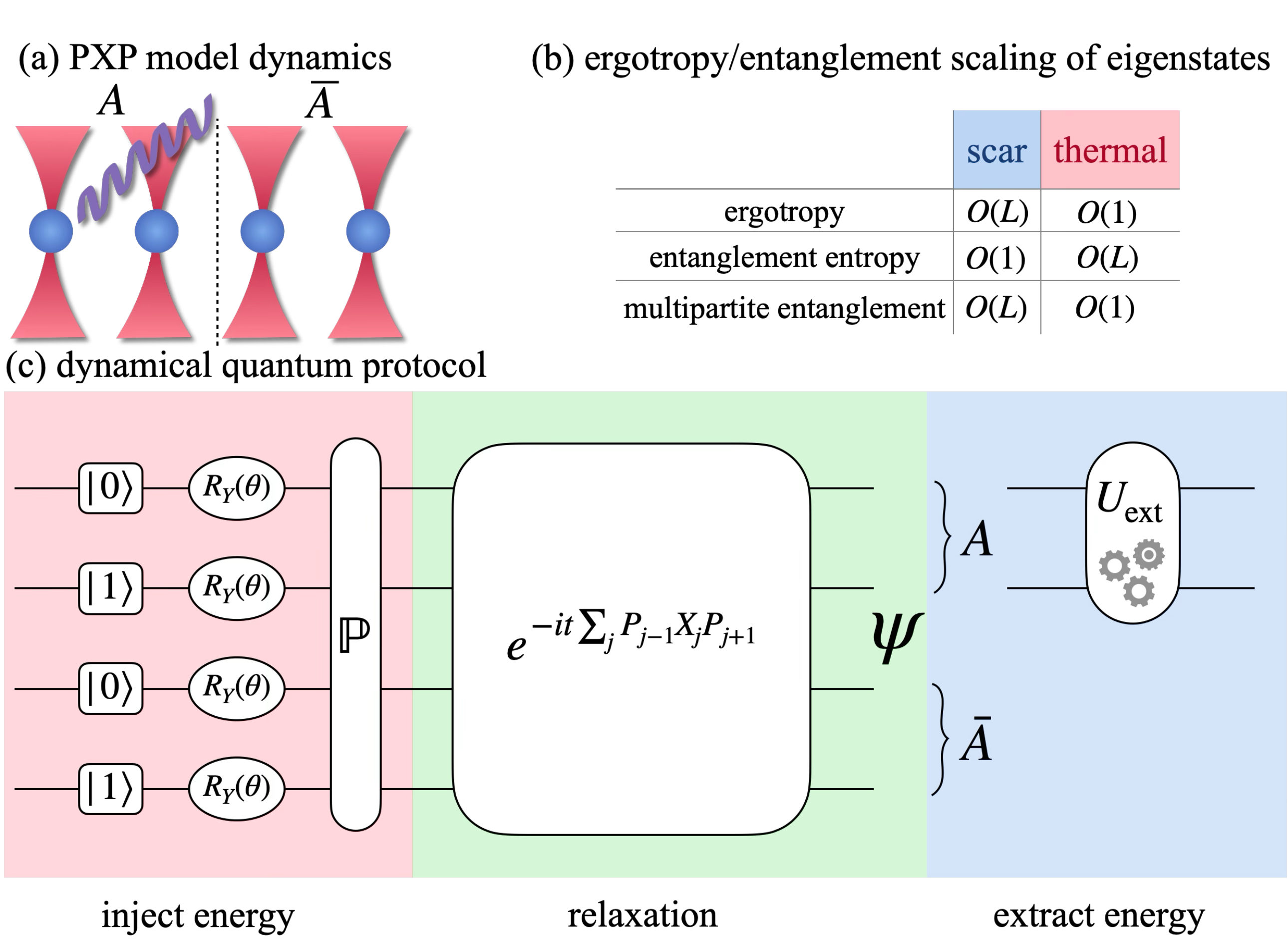}
    \caption{{\bf Schematics}. (a) {\bf PXP model dynamics}, realized in a blockaded Rydberg atom arrays with pulse that globally rotates the qubits. 
    (b) {\bf Ergotropy and entanglement scaling} for generic eigenstates. 
    (c) {\bf Dynamical quantum protocol} that uses reset together with a coherent rotation $R_Y(\theta) = e^{-i \frac{\theta}{2} Y}$ to inject energy, creating highly excited states for the PXP Hamiltonian. After relaxation, the subsystem of the steady state possesses extensive energy extractable by unitary operation. 
    }
    \label{fig:quantum_quench_protocol}
\end{figure}

\begin{figure*}[t!]
    \centering
    \begin{overpic}[width=0.335\textwidth]{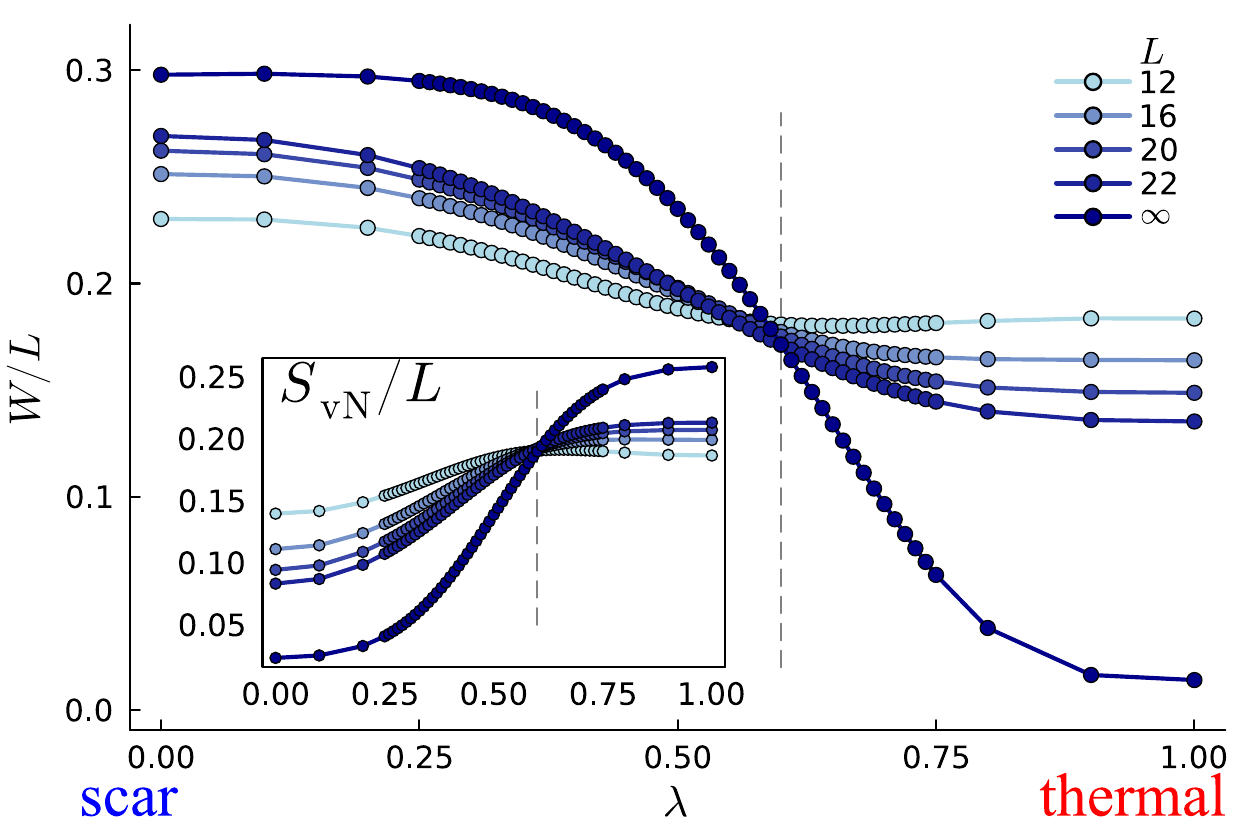}
        \put(-2,60){\textbf{(a)}}
    \end{overpic}
    \hfill
    \begin{overpic}[width=0.30\textwidth]{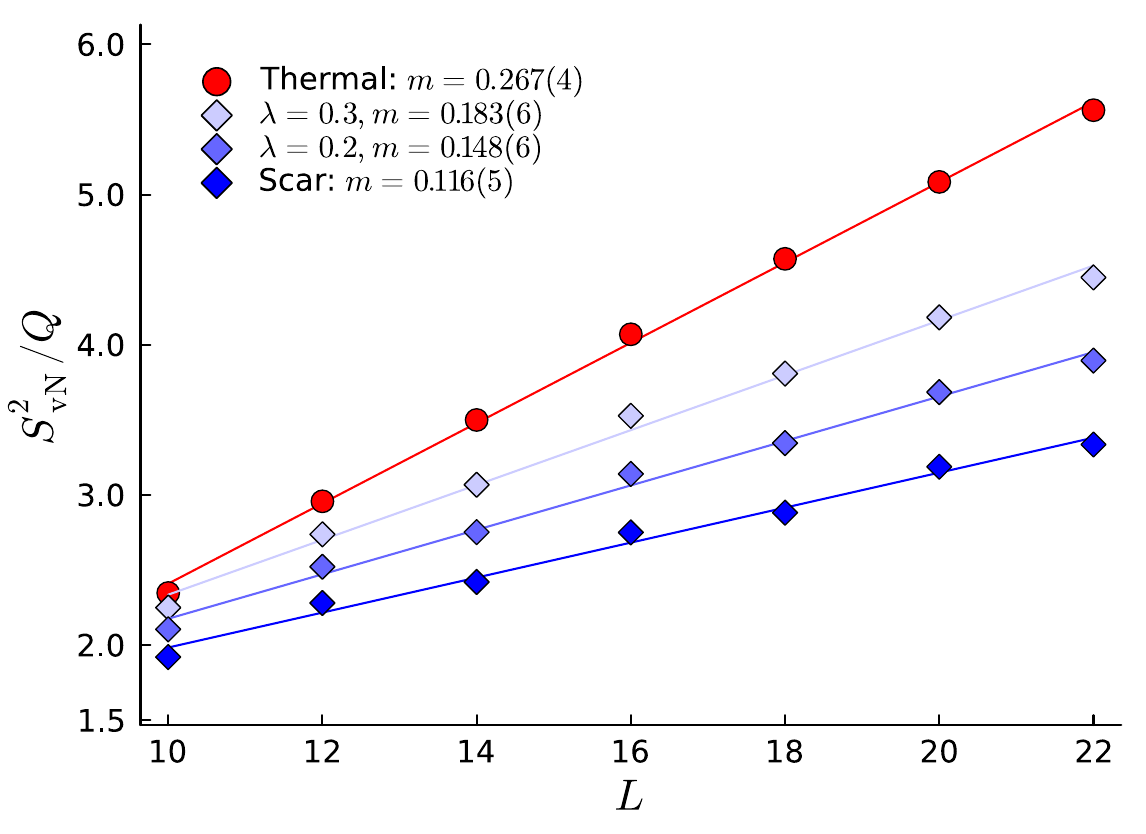}
        \put(-2,66.5){\textbf{(b)}}
    \end{overpic}
    \hfill
    \begin{overpic}[width=0.335\textwidth]{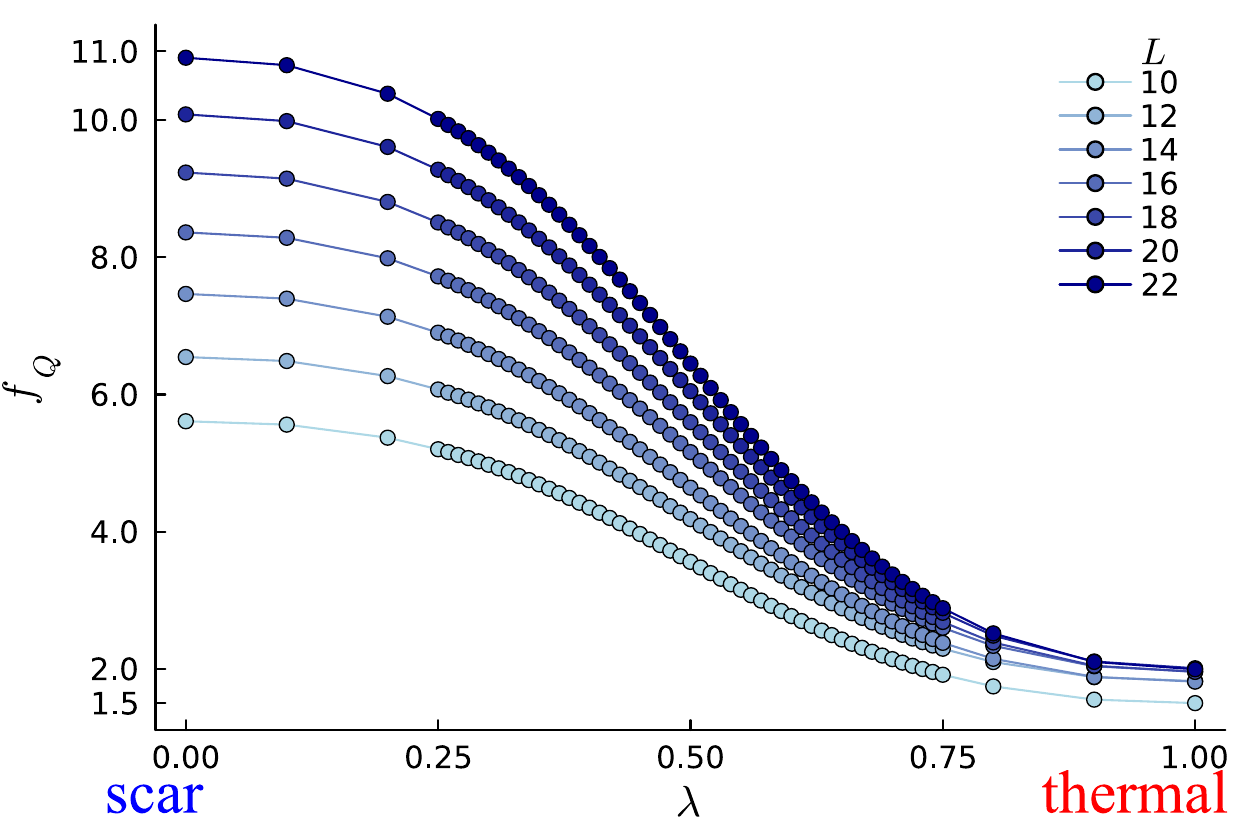}
        \put(-3,61){\textbf{(c)}}
    \end{overpic}
    \caption{
        {\bf Ergotropy and entanglement} for eigenstates of PXP model without restricting to specific symmetry sector by interpolating between scar and thermal states within the zero energy shell i.e. in the middle of the energy spectrum.
    {\bf (a)} {\bf Ergotropy density} $W/L$ exhibits a crossover from extensive to sub-extensive scaling. Data for $L=\infty$ is obtained by extrapolating finite size data to thermodynamic limit. The non-vanishing ergotropy at $\lambda=1$ reflects imperfect thermal state separation and finite size effects.
    Inset: Bipartite half chain entanglement entropy density $S_\mathrm{vN}/L$ showing transition from area law to volume law. We adopt phenomenological fitting forms: $S_{\mathrm{vN}}=a+ v L + c \ln L/3$, yielding thermodynamic limit ergotropy density $\lim_{L\to\infty} W/L$. Both observables show clear crossover around $\lambda_c \approx 0.6$. 
    {\bf (b)} {\bf Relation between entanglement entropy and bound energy}. Entanglement entropy squared $S_{\mathrm{vN}}^2$ versus bound energy $Q$ for system sizes $L=10$--$22$ shows a linear relation $S_{\mathrm{vN}}^2/Q = n+m L$, as an interacting many-body case generalizing the free fermion results in Refs.~\cite{Mula2023, Mitra_2025}. Each $\lambda$ is annotated with their slope $m$, and larger $\lambda$ (more thermal) yields larger slope $m$, indicating stronger suppression of ergotropy by entanglement. $m$ saturates quickly after $\lambda_c$ and thus only $\lambda = 0, 0.2, 0.3, 1$ are shown here. The fitting quality is depicted in supplemental material (SM)~\cref{fig:ergotropy_vs_entropy}(c).
    {\bf (c)} {\bf Multipartite entanglement} witnessed by quantum Fisher information (QFI) density $f_Q = \expval{( O-\langle O\rangle)^2}/L$ with respect to the antiferromagnetic operator $O = \sum_i (-1)^{i+1} Z_i$, showing a consistent crossover. Scar states, despite obeying area law entanglement entropy, possess extensive QFI density scaling $\propto L$, indicating genuine multipartite entanglement, in contrast to thermal states with sub-extensive QFI density~\cite{ExtensiveQFI_pappalardi}.
    }
    \label{fig:scarthermalErgo}
\end{figure*}

{\it Model Hamiltonian}.--
We focus on a one dimensional PXP chain with $L$ sites under periodic boundary conditions (PBC), which can be described by Hamiltonian:
\begin{equation}
H=\sum_{i=1}^L P_{i-1} X_{i} P_{i+1} 
\end{equation}
where site indices are identified as $i+L \equiv i$. Here $P=\ket{0}\bra{0}$ is the projector onto the Rydberg atom ground state and $X=\ket{0}\bra{1}+\ket{1}\bra{0}$ is the Pauli operator. The PXP model approximates the Rydberg atom arrays in experiment while capturing the essential correlated physics of the Rydberg blockade: two excited states are not allowed on nearest neighbours. The resulting effective Hilbert space dimension obeys the Fibonacci sequence and scales asymptotically with the Golden ratio $(\frac{\sqrt{5}+1}{2})^L$ when $L\gg 1$~\cite{Papic18reviewScar, Papic21reviewscar}. This model was found to host QMBS among its high energy eigenstates, with significant overlap with product states that are easily prepared in experiment, as manifested in quantum quench experiments. 
This model preserves translational symmetry $T$, inversion symmetry $I: i \rightarrow L-i+1$, and particle hole symmetry~\cite{Papic18reviewScar,Papic21reviewscar} defined by $\mathcal{P}=\prod_i Z_i$ with $\{\mathcal{P},H\}=0$, ensuring the energy spectrum is symmetric about zero energy. 
The Hilbert space $\mathcal{H}$ can be decomposed as:
\begin{equation*}
\mathcal{H} \approx\mathcal{H}_{\text{scar}}\oplus \mathcal{H}_{\text{thermal}}
\end{equation*} 
where the subspace hosting scar states $\mathcal{H}_{\text{scar}}$ could be described by the Forward Scattering Approximation (FSA)~\cite{Papic18reviewScar,EmergentSU2Scars,Papic21reviewscar}, a method that unveils the algebraic structure of a tower of nonthermal scar states, while $\mathcal{H}_{\text{thermal}}$ contains the finite density of thermal eigenstates.

The total many-body system is partitioned into two halves, $A$ and $\bar{A}$, each with $L/2$ qubits. We view $A$ as the target system for energy storage (i.e. ``battery''), and $\bar{A}$ plays the role of ``environment'' that interacts with $A$. The Hamiltonian can then be decomposed into $H = H_A + H_{\rm int} + H_{\bar{A}}$, where $H_A$ is the Hamiltonian of $A$ and $H_{\rm int}$ is the interaction between $A$ and $\bar{A}$. By tracing out $\bar{A}$, the target system $A$ lies in a mixed state in general: $\rho_A = {\rm tr}_{\bar{A}}\ketbra{\psi} $. 

{\it Entanglement and Ergotropy}.--
The stored energy in $A$ is a linear observable of the density matrix $E = \bra{\psi} H_A \ket{\psi} = {\rm tr}(H_A \rho_A)$. The interaction energy $H_{\text{int}}$ is omitted from such subsystem energy since we only consider work extraction from the subsystem $A$ itself.
However, for a mixed state, not all the excited energy is extractable by unitary operations, which is upper bounded by the ergotropy~\cite{ErgotropyOriginalAllahverdyan2004,Touil_2022_ergotropy_correlations,D'Alessio03052016,francica2020quantum}. Concretely, it refers to the maximal extractable work by applying unitary operations $U_{\rm ext}$ (supported on $A$ only):
\begin{equation}
W = E - \min_{U_{\rm ext}} \bra{\psi} U_{\rm ext}^\dag H_A U_{\rm ext} \ket{\psi}\equiv E - Q \ ,
\label{eq:ergotropy}
\end{equation}
where bound energy $Q\equiv \min_{U_{\rm ext}} \bra{\psi} U_{\rm ext}^\dag H_A U_{\rm ext} \ket{\psi}$ is interpreted as unextractable energy.
Note that $U_{\rm ext}$ is an extrinsic unitary operation, which should be distinguished from the intrinsic time evolution generated by Hamiltonian $H_A$. Since $U_{\rm ext}$ is designed to lower the energy of $A$, it usually does not commute with $H_A$.
The optimal unitary $U_{\rm ext}$ is a function of many-body reduced density matrix. To see this, first we go to the entanglement eigen-basis:
\begin{equation}
    \begin{split}
    \rho_A &\equiv e^{-H_{\rm ent}}
    = \sum_{n=1}^{{\rm dim} A} e^{-E_{{\rm ent},n}} \ketbra{E_{{\rm ent}, n}}  \ ,
    \end{split}
\end{equation}
with eigenvalues $p(n) = e^{-E_{{\rm ent},n}}$ that satisfy normalization condition $\sum_{n=1}^{{\rm dim} A} p(n) = 1$ with ${\rm dim}\ A$ being the Hilbert space dimension of subsystem $A$. 
As $\rho_A$ captures the bipartite entanglement between $A$ and $\bar{A}$, 
$H_{\rm ent}$ takes the physical meaning as the entanglement Hamiltonian, whose eigenvalues are dubbed as {\it entanglement} energy, sorted in ascending order: $E_{\rm ent, 1} < E_{\rm ent, 2} < \cdots $\footnote{If there is degeneracy in entanglement spectrum, reshuffling populations within a degenerate entanglement spectrum subspace leaves the ergotropy unchanged}. The distribution $\{ p(n) \}$ can be viewed as ``thermal'' distribution of energy $\{E_{\rm ent, n}\}$ at unit temperature. 
Unitary operation does not change the population distribution $\{ p(n) \}$, but instead changes the eigenstates. 
The optimal work extraction unitary maps these entanglement eigenstates to eigenstates of $H_A$ with energy eigenvalues $E_{n}$ in ascending order\cite{ErgotropyOriginalAllahverdyan2004}: 
\begin{equation}
    U_{\rm opt} (H_{\rm ent}, H_A)= \sum_{n=1}^{{\rm dim}A} |E_{n}\rangle \langle E_{{\rm ent},n}| \ .
    \label{eq:optimal_unitary}
\end{equation}
$U_{\rm opt} \rho_A U_{\rm opt}^\dag = \sum_{n=1}^{{\rm dim}A}e^{-E_{{\rm ent},n}} |E_{n}\rangle \langle E_{n}|$ is called {\it passive} state since its energy cannot be further lowered by any unitary operation\footnote{If the total system is in its ground state, the subsystem may not be in its ground state due to entanglement and correlations with the rest of the system}. Its energy determines the bound energy:
\begin{equation}
    \begin{split}
Q &
=  \bra{\psi} U_{\rm opt}^\dag H_A U_{\rm opt} \ket{\psi} = \sum_{n=1}^{{\rm dim} A} e^{-E_{{\rm ent},n}} E_{n} \ ,
\end{split}
\end{equation}
as a {\it non-linear} observable of the reduced density matrix $\rho_A$. 
It should be compared with another more well known non-linear observable, the von Neumann entropy that captures the bipartite entanglement between $A$ and $\bar{A}$:
\begin{equation}
S_{\rm vN} = -{\rm tr}\rho_A \ln \rho_A = \sum_{n=1}^{{\rm dim}A} e^{-E_{{\rm ent},n}} E_{{\rm ent},n} \ ,
\label{eq:entropy}
\end{equation}
which does not explicitly depend on $H_A$. In contrast, $U_{\rm opt}$ and $Q$ are functions of $H_{\rm ent}$ and $H_A$, i.e., the mismatch between the entanglement Hamiltonian and the physical Hamiltonian. 
Note that this is beyond the low energy phenomena described by the Li-Haldane conjecture that entanglement Hamiltonian could capture the physical Hamiltonian with boundary~\cite{LiHaldane2008}.
For systems that thermalize after long time evolution, $H_{\rm ent}\propto H_A$~\cite{Huse_MBL_ETH,D'Alessio03052016,Thermalization_Deutsch_1991,Thermalization_Rigol_2008,QD_Canonical_Tasaki1998,Thermalization_Srednicki_1994}, the optimal unitary approximates the identity operator, and $W$ approaches zero. 
Intuitively, higher entanglement corresponds to bound energy, thereby suppressing ergotropy~\cite{Mula2023,Mitra_2025,Yang22coherence,Touil_2022_ergotropy_correlations}. The quantum many-body scar violates thermalization and can potentially show higher ergotropy.

{\it Eigenstates ergotropy}.--
Within the same energy shell, the ergotropy of the eigenstates can differ significantly between the thermal states and the scar states, owing to their distinctive entanglement spectra. 
We consider an ensemble of pure states interpolated between the scar and the thermal states, which is found to exhibit a crossover of the ergotropy from extensive to sub-extensive scaling behavior. 
Concretely, to eliminate the energy dependence of ergotropy across different energy eigenstates, we consider the $E=0$ energy subspace (corresponding to the infinite temperature limit and the typicality regime), where there is an ensemble of pure thermal states $\ket{\rm thermal}_n$ with index $n$ labeling each one, and a pure scar state $\ket{\rm scar}$~\footnote{Although there may exist 2 or more $E=0$ scars as suggested in Refs.~\cite{number_scar, ivanov2025exactarealawscareigenstates}, we only use one scar described by FSA, which is inversion symmetric and translational symmetric: $I\ket{{\rm scar}} = \ket{{\rm scar}}, T\ket{{\rm scar}} = \ket{{\rm scar}}$}. To isolate the scar component from hybridized thermal states, we construct it by diagonalizing $P_{E=0}P_{\text{FSA}}P_{E=0}$, where $P_{E=0}$ and $P_{\text{FSA}}$ project onto the zero energy and FSA subspaces respectively (details provided in SM).

We investigate the ergotropy \cref{eq:ergotropy} and entanglement entropy \cref{eq:entropy} by partitioning the following state into two halves:
\begin{equation}\label{state_zero_energy}
\ket{\psi_n(\lambda)} = (1 -\lambda) \ket{\rm scar} + \lambda \ket{\rm thermal}_n \ .
\end{equation}
Note crucially that the thermal states on the same energy shell form an ensemble. Since the scar state resides in the inversion-symmetric sector ($I=1$), we project all thermal states onto the inversion symmetric sector before superposition and averaging over their contributions, ensuring that no symmetry-sector mixing affects the discrimination between thermal and scar states.
As shown in \cref{fig:scarthermalErgo}(a) inset, the system undergoes a crossover from scar to thermal, witnessed by an entanglement transition from the known sub-volume law $S_{\mathrm{vN}} \sim \ln L$ for scar state~\cite{Papic18reviewScar}, to volume law $S_{\mathrm{vN}} \sim L$ for thermal state. In comparison, ergotropy exhibits a crossover from extensive scaling for the scar state, to sub-extensive scaling for the thermal state. To demonstrate that these results do not depend on the large degeneracy of zero-energy modes in the PXP model, we have extended the analysis to the finite energy shell $[E-\Delta E, E+\Delta E]$, where the extensive degeneracy is lifted and only a few symmetry resolved states reside in the narrow window, see End Matter~\cref{fig:new_scar_scaling_symmetry}.
\begin{figure*}[htpb!] 
    \centering
    \begin{overpic}[width=0.325\textwidth]{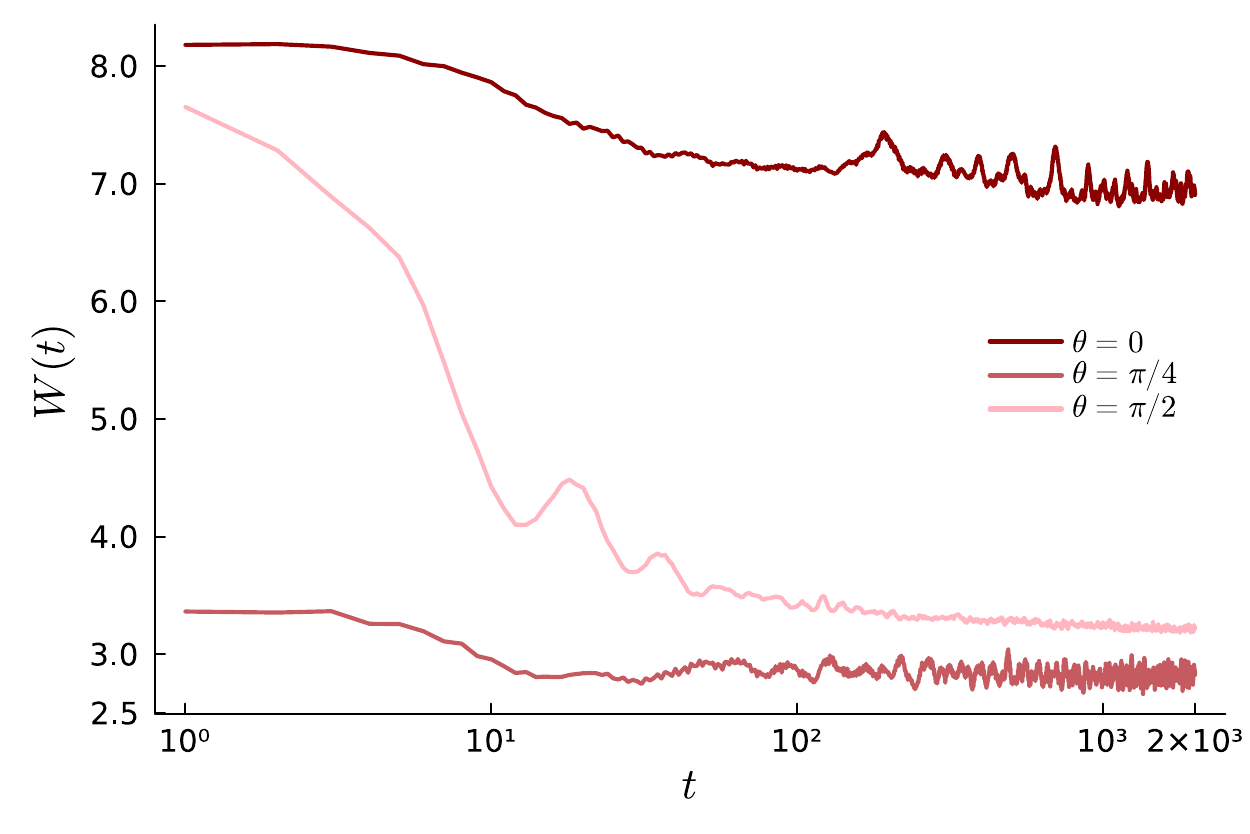}
        \put(-1, 60){\textbf{(a)}}
    \end{overpic}
    \begin{overpic}[width=0.325\textwidth]{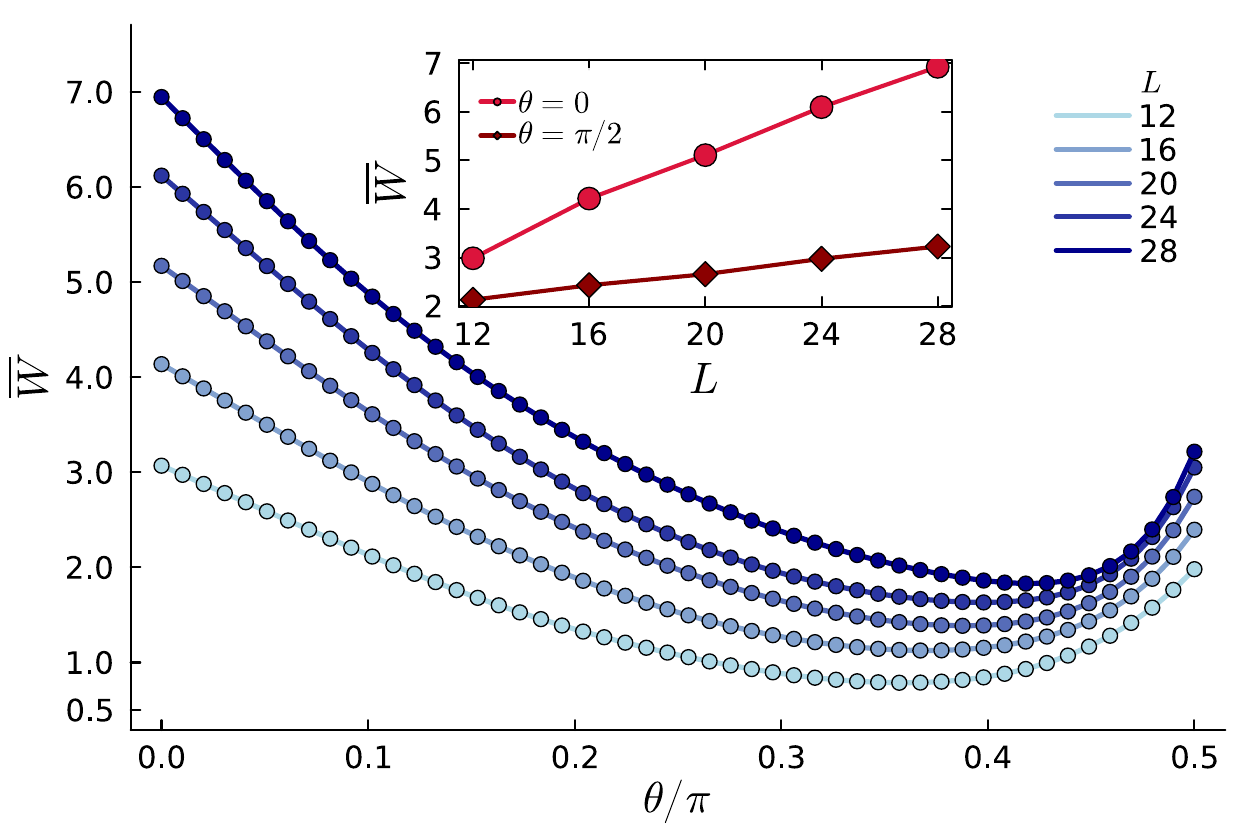}
        \put(-2, 60){\textbf{(b)}}
    \end{overpic}
    \begin{overpic}[width=0.325\textwidth]{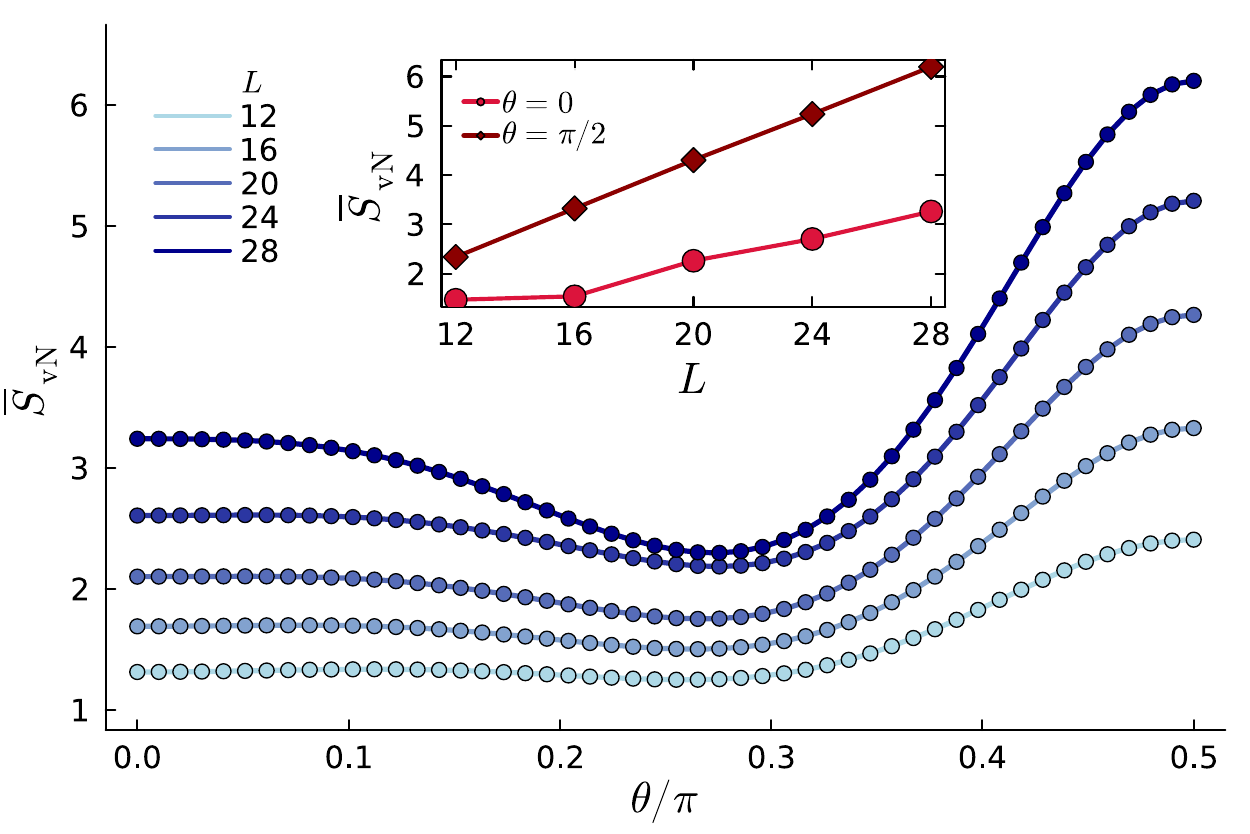}
        \put(-2, 60){\textbf{(c)}}
    \end{overpic}
    \caption{\textbf{Quantum quench dynamics and steady state properties.} 
    \textbf{(a)} {\bf Time evolution of ergotropy} $W(t)$ for $L=28$ reveals distinct relaxation patterns across rotation angles ($\theta = 0, \pi/4, \pi/2$). Scarred dynamics ($\theta=0$) exhibit persistent oscillations with revivals, while thermal dynamics ($\theta=\pi/2$) show rapid decay. The intermediate case ($\theta=\pi/4$) displays scar like behavior with reduced initial ergotropy due to lower injected energy (cf. SM \cref{eq:energy_density}). 
    \textbf{(b)} {\bf Steady state ergotropy} $\bar{W}$ exhibits non-monotonic $\theta$ dependence with extensive scaling for scar and sub-extensive scaling for thermal limit across system sizes $L=12$--$28$. The minimum approaches thermal regime in thermodynamic limit, suggesting the thermodynamic advantage of scarred states.
    \textbf{(c)} {\bf Steady state entanglement entropy} $\bar{S}_{\mathrm{vN}}$ shows behavior inversely related to ergotropy, transiting from low entanglement scar regime to volume law thermal regime, demonstrating fundamental entanglement-ergotropy anti-correlation. The detailed dynamics of bound energy $Q(t)$ and entanglement entropy $S_{\mathrm{vN}}(t)$ is presented in SM \cref{fig:additional_dynamics}.
    }
    \label{fig:quench_dynamics}
\end{figure*}

In fact, the anti-correlation between ergotropy and entanglement manifests in a phenomenological relationship $S_{\mathrm{vN}}^2/Q = n+m L$, which was derived for  non-interacting fermion chain in~\cite{Mula2023,Mitra_2025}, and is numerically verified for the {\it interacting} PXP model in our case, see \cref{fig:scarthermalErgo}(b). 

To corroborate the anti-correlation between ergotropy of $A$ and the entanglement of the total system $A\cup \bar{A}$, we also investigate the multipartite entanglement witnessed by the quantum Fisher information (QFI)~\cite{QFI_ME,ME_Metro}, which is distinct from the bipartite entanglement entropy. For example, it was shown that volume law entangled states possess little QFI, while scar states are rich in QFI~\cite{ExtensiveQFI_pappalardi}. 
Here by tuning the angle between scar and thermal states, we observe that QFI density crosses over from extensive to sub-extensive, which is consistent with the phenomena observed in Ref.~\cite{ExtensiveQFI_pappalardi}. 
In addition, bipartite and tripartite mutual information are shown in SM \cref{fig:mutualinfofisherinfo}. 
Overall, these entanglement measures witness the strong anti-correlation between ergotropy and entanglement among eigenstates of the Hamiltonian.


{\it Quantum quench dynamics with coherent rotation}.--
To harness the $E=0$ eigenstates, we consider a quantum quench~\cite{Bernien2017} that suddenly resets the system into a product state $\ket{\mathbb{Z}_2}\equiv \ket{1010\cdots10}$ lying in the highly excited space with $\bra{\mathbb{Z}_2}H\ket{\mathbb{Z}_2} = 0$, which injects finite energy density into the system. This state is then evolved under the intrinsic PXP Hamiltonian. At late times, the resulting evolved state is expected to be the superposition of scar and thermal eigenstates, where the scar states (in particular, the scar eigenstate at $E=0$) have the largest amplitude as found in Ref.~\cite{Papic18reviewScar}.
We can tune the superposition by employing a coherent rotation for $\ket{\mathbb{Z}_2}$, as shown in \cref{fig:quantum_quench_protocol}(c), 
characterized by a uniform angle $\theta\in[0,\pi/2]$ for every qubit: $e^{-i\frac{\theta}{2} \sum_{i=1}^L Y_i}$ where $Y_i$ denotes the Pauli $Y$ operator on qubit $i$. 
This rotation continuously tunes the overlap of the system state with the scar states. Implementing the Rydberg blockade constraint, we perform $\mathbb{P} e^{-i\frac{\theta}{2} \sum_{i=1}^L Y_i} \mathbb{P}$ to restrict the rotation to act on the physical space, where $\mathbb{P} = \prod_{i}^{L} \frac{1}{2}(1+{\rm CZ}_{i,i+1})$, with ${\rm CZ}$ being the control Z matrix. 
The quantum state follows as
\begin{equation}\label{state}
\ket{\psi(\theta)} =\prod_i \frac{1+{\rm CZ}_{i,i+1}}{2} e^{-i\frac{\theta}{2} \sum_i Y_i} \ket{1010\cdots} \ ,
\end{equation}
which admits a bond dimension 2 matrix product operator representation:
\begin{equation}\label{eq:state_tensor}
    \includegraphics[width=.8\columnwidth]{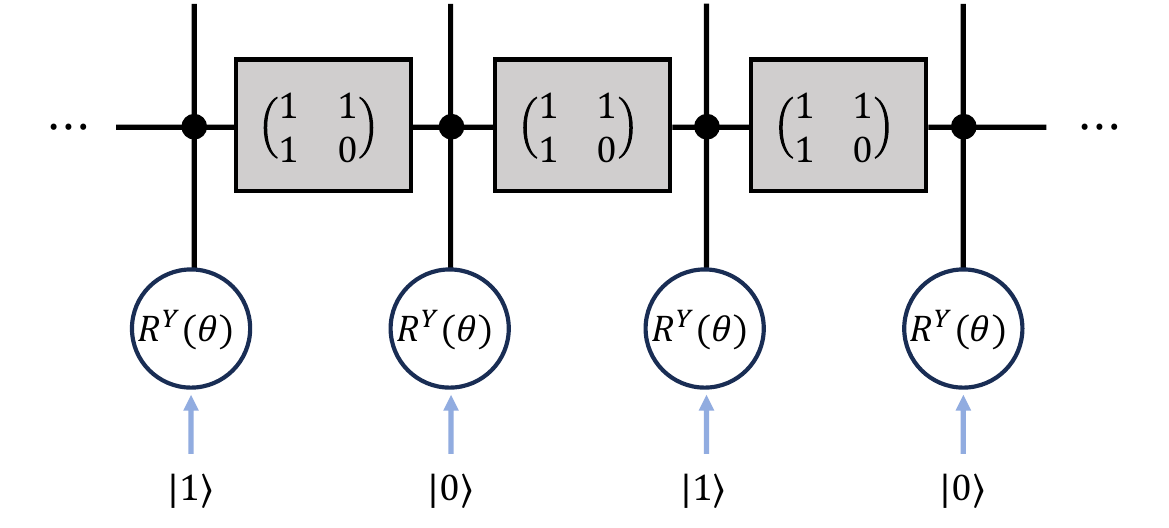} \ ,
\end{equation}
where the black dot refers to rank 4 delta tensor with bond dimension 2. And such MPO have $\ket{\mathbb{Z}_2}$ as input. There are two plausible ways to prepare \cref{eq:state_tensor} state: (i) the analogue route, adiabatically preparing this state by starting from the product state and slowly turning on the Rydberg interaction to remove the blockaded excitations; (ii) the digital route, where the bond dimension 2 MPS can be prepared by a staircase unitary circuit with $L$ gates, with each two body gate fixed by the left canonical form of the local tensor~\cite{Schon2005sequential}.
The subleading eigenvalue of the normalized transfer matrix is $\lambda = \frac{2\cos(2\theta)-f(\theta)+14}{2\cos(2\theta)+f(\theta)+14}$, where $f(\theta) = \sqrt{2} \sqrt{44 \cos(2\theta) - 3 \cos(4\theta) + 87}$, which leads to finite correlation length $\xi = -1/\ln\lambda$. 
When an infinite long open chain is partitioned into two halves with a single entanglement edge, the two entanglement eigenvalues are $\frac{1}{2} \pm  \sqrt{\frac{1}{4}-h(\theta)}$, where $h(\theta) = \frac{128\sin^6\theta}{(2\cos(2\theta) + f(\theta)+14)f^2(\theta)} \leq \frac{2}{15+5\sqrt{5}} < \frac{1}{4}$, meaning that it's not a Haldane phase. It is symmetric around $\theta=\pi/2$ i.e. with the same entanglement spectrum between $\theta$ and $\pi-\theta$, related by a global flip $\prod_j X_j$. 

Such low entangled states possess extensive subsystem ergotropy, depending on $\theta$. Their relaxation dynamics under the intrinsic PXP Hamiltonian evolution $U = \mathrm{e}^{-iHt}$ is revealed in the following. Firstly, the interaction between $A$ and $\bar{A}$ develops entanglement between them, while mixing the states in $A$. The spread of entanglement raises the bound energy (see SM ~\cref{fig:additional_dynamics}), in accordance with the physical picture established for the eigenstates. Consequently, $\rho_A(t)$ relaxes towards its passive state as time evolves. Secondly, we notice that the subsystem energy is conserved: $\bra{\phi(\theta)}\mathbb{P} \comm{H_A}{H_{{\rm int}}}\mathbb{P}\ket{\phi(\theta)} = 0$ (cf. SM \cref{Equ:sub_energy_conservation}).
We define the time-dependent ergotropy $W(t)$ and bound energy $Q(t)$ as those of the reduced state $\rho_A(t)$ at time $t$ during the dynamics. Due to the conservation of the subsystem energy, the ergotropy dynamics $W(t)=E-Q(t)$ is determined by the bound energy dynamics $Q(t)$, which decays upon time evolution to the saturated value of the steady state. 

We analyze the ergotropy in comparison with the entanglement~\cite{turner2018weak} of $\ket{\psi(\theta,t)}=U(t) \ket{\psi (\theta)}$ as shown in \cref{fig:quantum_quench_protocol} in the zero momentum, inversion symmetric sector. The system sizes are chosen to be $L=4N$ where $N$ is an integer, so as to avoid even odd effects of the subsystem energy. Ergotropy dynamics exhibits three representative quench behaviors as shown in \cref{fig:quench_dynamics}(a): 
For $\theta=0$ (with maximal scar overlap), $W(t)$ decays slowly with persistent late time revivals, maintaining sizable extractable work at late times as a hallmark of non-thermal dynamics. The intermediate case $\theta = \pi/4$ exhibits scar like oscillations and relaxation, with relatively small initial ergotropy due to its lower initial energy (cf. SM \cref{eq:energy_density}). In contrast, the $\theta=\pi/2$ quench shows rapid ergotropy suppression, finally damping to a small steady value, akin to thermal state behavior. The entanglement growth dynamics is provided in SM \cref{fig:additional_dynamics}.
These dynamics reveal an anti-correlation between entanglement and ergotropy: the growth of $S_{\mathrm{vN}}(t)$ is accompanied by the suppression of $W(t)$.

The steady state ergotropy and entanglement entropy for generically tuned angle $\theta$ are shown in \cref{fig:quench_dynamics}(b,c).
The ergotropy $\overline{W}(\theta)$ shows a non-monotonic dependence on $\theta$, which could be a finite size effect, as the minimum point shifts towards the $\theta=\pi/2$ limit upon increasing the system size. The entanglement entropy, on the other hand, exhibits a more obvious non-monotonic behavior with tuning $\theta$. Note that near $\theta=0$, the dynamics is strongly fluctuating due to the scar contribution, see SM~\cref{fig:additional_dynamics}. Our results indicate that $\theta=0$ is indeed the optimal angle for the energy storage and extraction.

The ergotropy as the optimal extractable work requires a unitary operation $U_{\rm opt}$~\eqref{eq:optimal_unitary}. Here we compile it approximately with a variational translation invariant unitary circuit of low depth. The shallow circuit is composed of uniformly-addressed ZYZ rotations and nearest-neighbor CZ gates (see End Matter \cref{fig:compiled_circuit_depth}).
With only depth $D=1$ i.e. uniform global rotation, the circuit already extracts substantial fraction ($\gtrsim 50\%$) of work for the time evolved scar state at $\theta=0$, while increasing depth $D$ systematically lifts the work curve towards the ergotropy bound.
These results demonstrate that the ergotropy signal is not merely a theoretical construction but is accessible with hardware-efficient operations native to Rydberg atom arrays and superconducting processors.

{\it Discussion and outlook}.--
The ergotropy dichotomy between thermal and scar states originates from their contrasting entanglement structures: volume-law entanglement constrains extractable work to be sub-extensive, whereas area-law entanglement permits extensive ergotropy. Such connection can be potentially generalized to other ergodicity breaking systems beyond PXP model, such as the variety of systems hosting scars~\cite{Regnault22reviewscar, Papic21reviewscar,Lin2020_2D_scar,SGA_scar_Moudgalya2020,scars_structure} or many-body localization~\cite{Pal2010_MBL,Abanin2019_RMP_MBL,Lukin_2019_MBL_Entanglement,Huse_MBL_ETH}. Among them, Rydberg atom arrays have emerged as a particularly versatile platform for observing and manipulating quantum many-body scars~\cite{bluvstein2021controlling,Bernien2017,liang2025observation}, which natively favor uniform global rotation that is crucial in our protocol and continuously reloading fresh atoms~\cite{Bluvstein_2022_processor,Bluvstein_2023_reconfigurable_processor}. 
Moreover, quantum many-body scar states have recently been engineered on a superconducting qubit processor~\cite{zhang2023many}, which allows us to compile the unitary operator for energy extraction. Last but not least, provided the advanced quantum technology of mid-circuit measurement and feed-forward operations in intermediate scale quantum devices~\cite{ryananderson2021realizationrealtimefaulttolerantquantum,Bluvstein_2022_processor}, and the rapid development of the open quantum systems dynamics~\cite{Fisher2022reviewMIPT, Potter21review, NishimoriCat, Chen25nishimori, Zhu24floquetcode, teleportcode, Puetz25percolation, Wang25selfdual}, one does not have to be constrained within a unitary approach. 
While unitary operation only extracts energy from the isolated system, quantum measurement can extract information to generalize the Maxwell's demon~\cite{Bennett_1987_demon,leff1990maxwell} and Szilard engine~\cite{Szilard1929,Landauer_1961} from few-body experiments~\cite{Toyabe2010, Koski_2014_Szilard_engine,Peterson_2016_Landauer} to quantum many-body system, for a more versatile control of the energy transport and information processing.  

{\it Code and data availability}.-- 
The numerical data shown in the figures is available on Zenodo~\cite{zenodo}. The numerical code is accessible on GitHub~\cite{github}. \\

{\it Acknowledgement}.--
GYZ acknowledges the support of National Natural Science Foundation of China - Young Scientists Fund (grant no.~12504181), Start-up Fund of HKUST(GZ) (grant no.~G0101000221), Guangdong provincial project (grant no.~2024QN11X201) and Guangdong Basic and Applied Basic Research Foundation (grant no.~2026A1515010965). JGL acknowledges the support of the National Natural Science Foundation of China under grant nos. 12404568.

\bibliography{ergotropy_ref}

\widetext
\clearpage
\appendix
\section{End Matter}
\setlength{\textfloatsep}{1.0em}
\setlength{\floatsep}{1.0em}
\setlength{\intextsep}{1.0em}
\setlength{\abovecaptionskip}{0.6em}

\subsection{Compile unitary operator $U_{\rm opt}$ into shallow quantum circuits}

\begin{figure}[H]
    \centering
    \raisebox{0.006\textwidth}{%
    \begin{overpic}[height=0.208\textwidth]{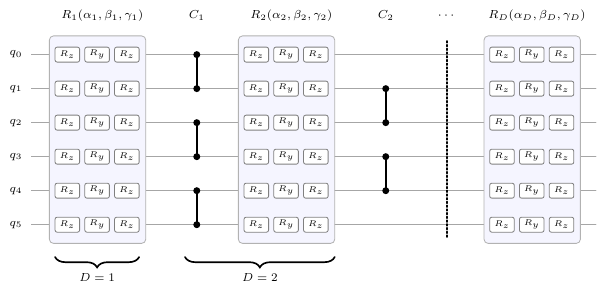}
        \put(-6,48){\textbf{(a)}}
    \end{overpic}}\hspace{0.002\linewidth}
    \begin{overpic}[height=0.208\textwidth]{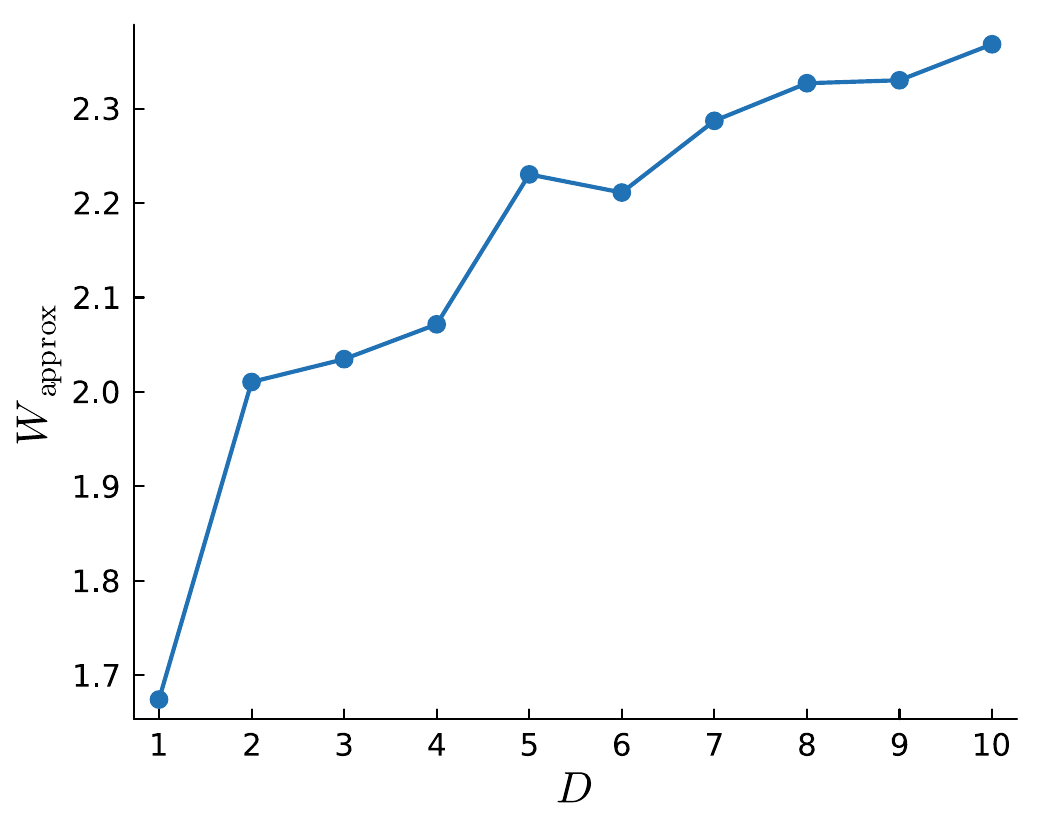}
        \put(-6,83){\textbf{(b)}}
    \end{overpic}\hspace{0.002\linewidth}
    \begin{overpic}[height=0.208\textwidth]{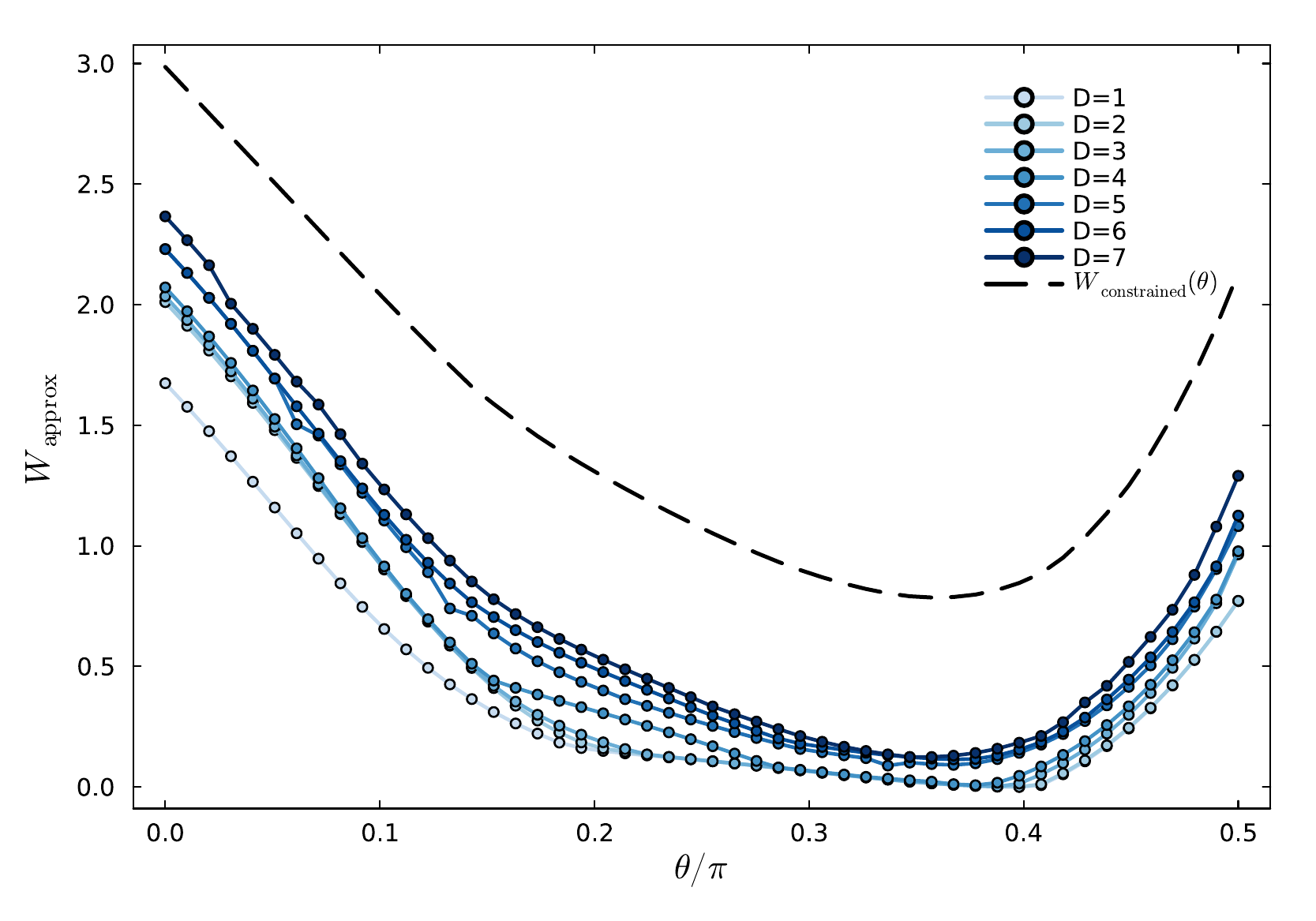}
        \put(-6,73){\textbf{(c)}}
    \end{overpic}
    \caption{\textbf{Circuit ansatz and depth scaling.}
    \textbf{(a)} Globally addressed ansatz used to compile the ideal constrained-space extraction unitary.
    Each rotation layer applies the same ZYZ Euler rotation $r_d=R_z(\alpha_d)R_y(\beta_d)R_z(\gamma_d)$ to all sites, where $d=1,\ldots,D$, and the entangling layer is CZ gates.
    The circuit acts on the full qubit Hilbert space and may leave the blockade subspace during the pulse sequence, the constraint is applied at the input and output of the circuit and the PXP operation is evaluated as $K_D=V^\dagger U_D V$.
    \textbf{(b)} Optimized $W_{\rm approx}$ versus repetition layer depth $D$ for the scarred initial state at $L=12,l=6$.
    The extracted work of the compiled unitary grows from about $1.67$ at $D=1$ to about $2.36$ at $D=10$.
    \textbf{(c)} Optimized $W_{\rm approx}$ as a function of the input rotation angle $\theta$ at $L=12$, overlaid for repetition layer depths $D=1,2,\ldots,7$.
    The colored curves show the optimized approximate work for each repetition layer depth, while the black dashed curve denotes the ergotropy $W_{\rm constrained}(\theta)$.
    Increasing repetition layer depth $D$ systematically brings $W_{\rm approx}$ closer to $W_{\rm constrained}(\theta)$.}
    \label{fig:compiled_circuit_depth}
\end{figure}

Here we provide a concrete example of compiling the optimal unitary operator $U_{\rm opt}$ for work extraction into a shallow quantum circuit, which can be implemented on near-term quantum devices.

Motivated by Rydberg atom arrays, where global laser pulses naturally implement spatially uniform operations, we use a globally addressed ansatz: each single qubit layer applies the uniform ZYZ Euler rotation to all sites, and the entangling layers are applied nearest-neighbor CZ gates in a brick-wall pattern.
Throughout this subsection, the repetition layer depth $D$ means the number of uniform ZYZ rotation layers.
With this convention, $D=1$ is a purely uniform ZYZ rotation circuit containing only $R_1$.
The next depth, $D=2$, appends the pair $C_1$ and $R_2$ in circuit order, and each further increase of $D$ adds one more CZ layer followed by one global rotation layer.
Explicitly,
\begin{equation}
    \begin{alignedat}{2}
    U_1(\boldsymbol{\vartheta})
    &=R_1,\qquad&
    R_d
    &=
    r_d\otimes r_d\otimes\cdots\otimes r_d,\\
    U_D(\boldsymbol{\vartheta})
    &=R_D C_{D-1} R_{D-1}\cdots C_1 R_1,\quad D\geq 2,\qquad&
    r_d
    &=R_z(\alpha_d)R_y(\beta_d)R_z(\gamma_d)
    =
    e^{-i\alpha_d Z/2}e^{-i\beta_d Y/2}e^{-i\gamma_d Z/2}.
    \end{alignedat}
\end{equation}
where the tensor product contains $l$ identical single-qubit rotations, so the $d$th rotation layer shares one parameter set $(\alpha_d,\beta_d,\gamma_d)$ across all sites.
Here $C_d$ denotes the $d$th brick-wall CZ layer, with $d=1,\ldots,D-1$.
Thus a depth-$D$ ansatz contains $D$ uniform ZYZ rotation layers and $D-1$ CZ entangling layers, giving only $3D$ variational parameters, independent of the subsystem size.

The unitary $U_{\rm opt}$ is the ideal extraction map in the constrained Hilbert space of the PXP subsystem.
It gives the ergotropy upper bound by rearranging the eigenvectors of $\rho_A$ according to the energy ordering of $H_A$.
To connect this bound with experiment, we use a variational quantum circuit compilation approach to approximate the ideal ergotropy unitary with a shallow ansatz on the full qubit Hilbert space, directly suitable for platforms such as Rydberg atom arrays and superconducting processors.
We optimize this circuit by the extracted work, rather than by gate fidelity, so that the variational unitary implements the thermodynamic task itself.

The physical picture is the following.
Let $V:\mathcal{H}_{\rm c}\to\mathcal{H}_{\rm full}$ be the isometry embedding the constrained basis into the full $2^l$-dimensional qubit Hilbert space.
The experimental circuit acts in $\mathcal{H}_{\rm full}$ and is allowed to leave the constrained subspace at intermediate times.
The effective operation relevant for the PXP battery is then obtained by projecting the final state back through the isometry to obtain the accessible work
\begin{equation}
    \begin{aligned}
        K_D(\boldsymbol{\vartheta}) &= V^\dagger U_D(\boldsymbol{\vartheta}) V , \\
    W_{\rm approx}(\boldsymbol{\vartheta})
    =
    \tr(\rho_A H_A)
    & -
    \tr\!\left[
    H_AK_D(\boldsymbol{\vartheta})\rho_AK_D^\dagger(\boldsymbol{\vartheta})
    \right].
\end{aligned}
\end{equation}
For each $\rho_A(\theta)$ and fixed $D$, we optimize
\begin{equation}
    \boldsymbol{\vartheta}_{\rm opt}(\theta,D)=\operatorname*{arg\,max}_{\boldsymbol{\vartheta}=\{\alpha_d,\beta_d,\gamma_d\}_{d=1}^{D}\in\mathbb{R}^{3D}} W_{\rm approx}(\boldsymbol{\vartheta};\theta,D).
\end{equation}
Numerically, we perform the maximization using multistart L-BFGS~\cite{liu1989limited,liu2018differentiable} and retain the run with the largest $W_{\rm approx}$.
This avoids an artificial fidelity target, since unitaries differing within degenerate eigenspaces can extract the same work.

\begin{figure*}[!htbp]
    \centering
    \begin{minipage}[t]{0.3\textwidth}
        \begin{overpic}[width=\linewidth]{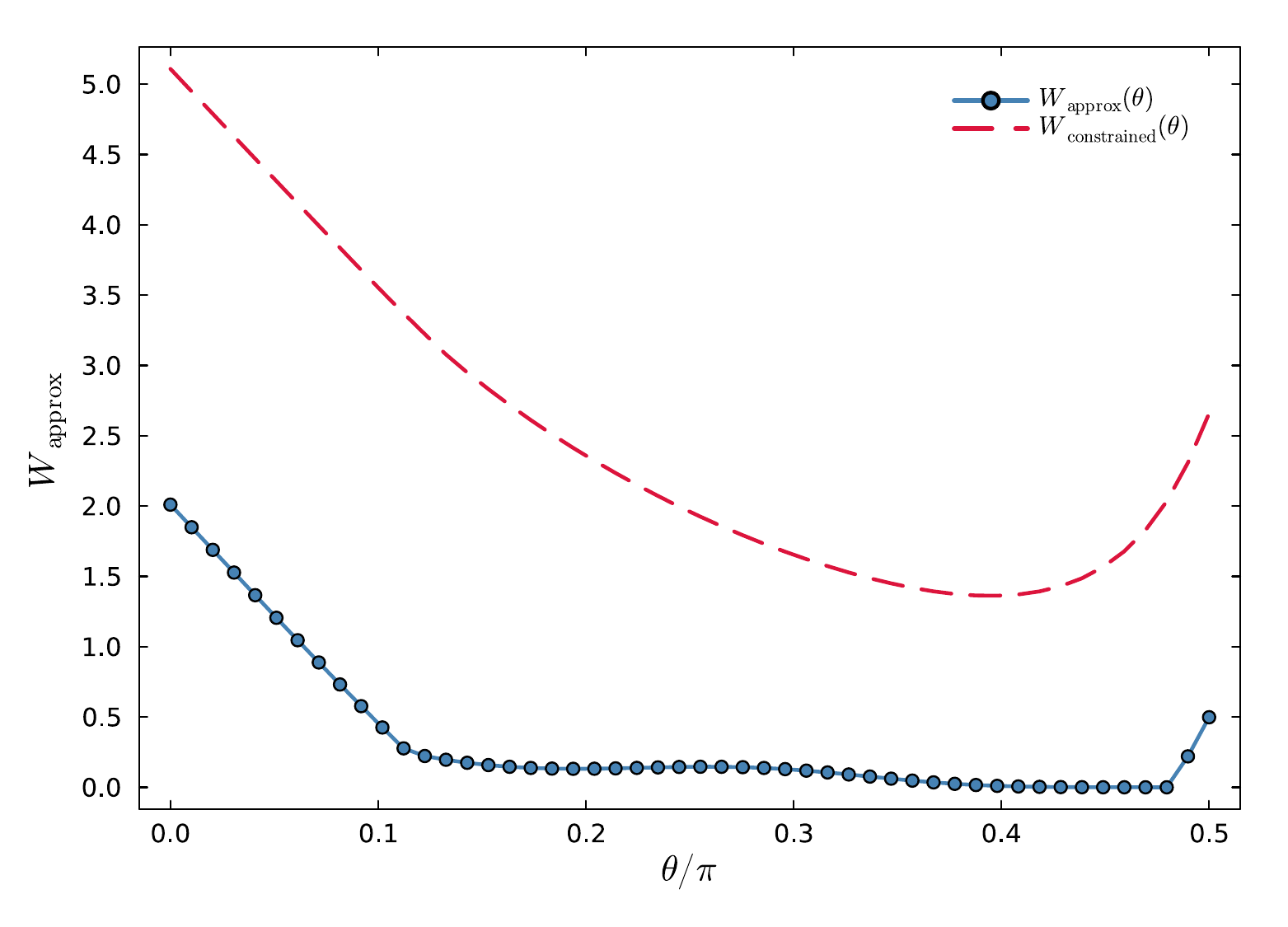}
            \put(-2,60){\textbf{(a)}}
        \end{overpic}
    \end{minipage}\hspace{0.01\textwidth}
    \begin{minipage}[t]{0.3\textwidth}
        \begin{overpic}[width=\linewidth]{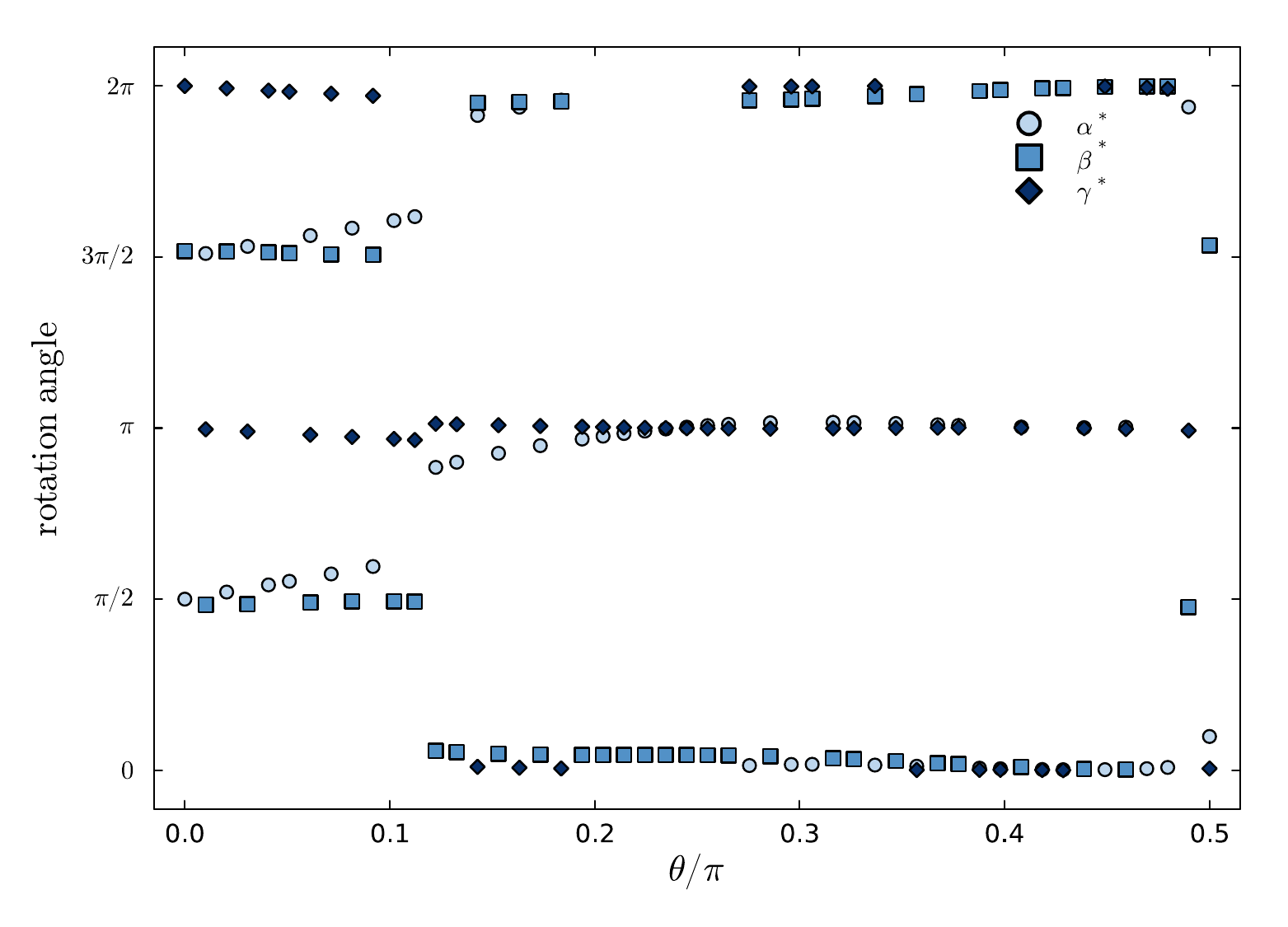}
            \put(-2,60){\textbf{(b)}}
        \end{overpic}
    \end{minipage}\hspace{0.01\textwidth}
    \begin{minipage}[t]{0.23\textwidth}
        \raisebox{0.004\textwidth}{%
        \begin{overpic}[width=\linewidth]{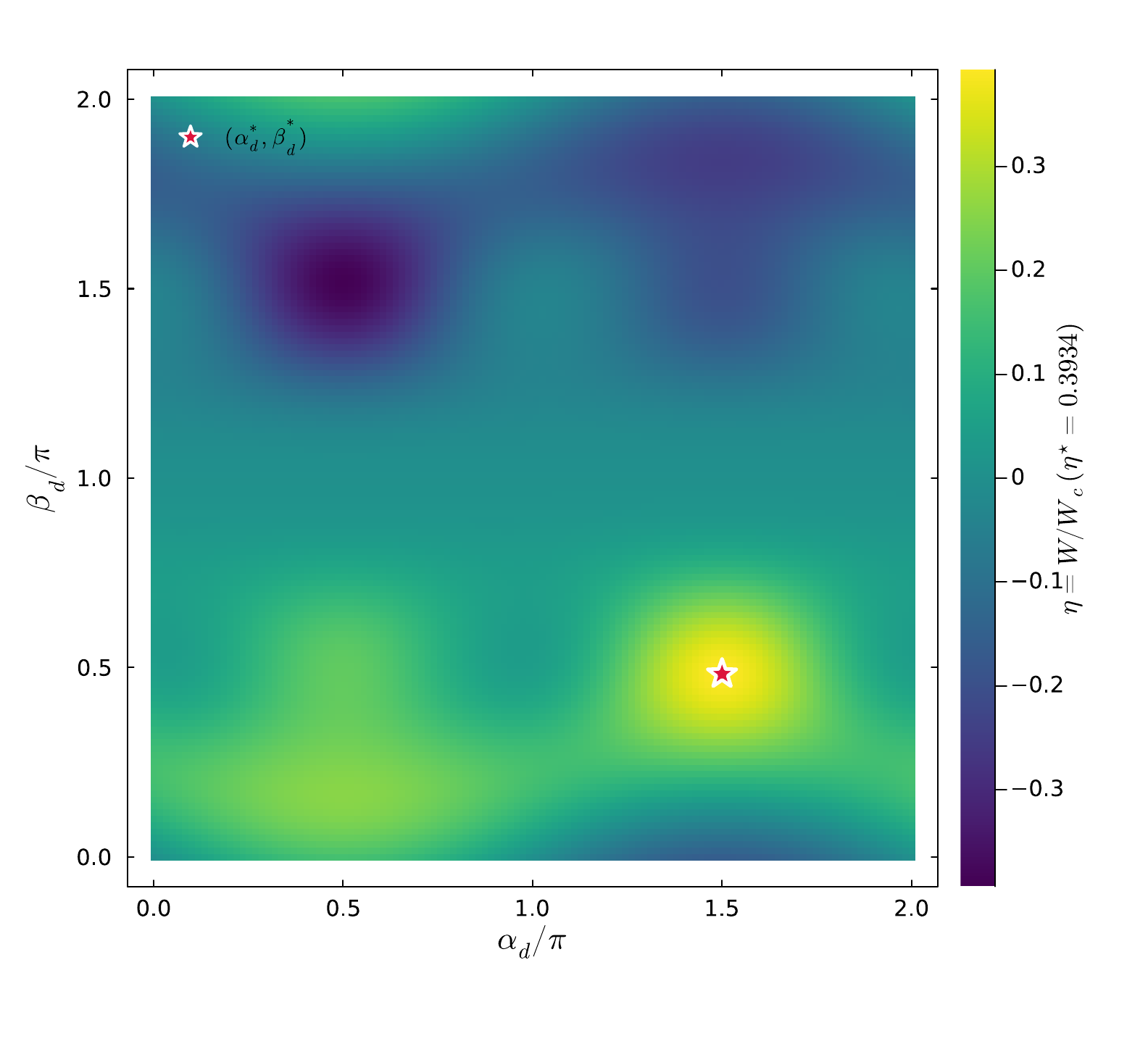}
            \put(-2.5,80){\textbf{(c)}}
        \end{overpic}}
    \end{minipage}
    \caption{\textbf{Angle dependence and variational landscape of the compiled extraction circuit.}
    \textbf{(a)} Optimization of the extraction quantum circuit for the PXP quench family at $L=20$ and $D=1$.
    For each $\theta$, the ZYZ pulse angles are optimized to maximize $W_{\rm approx}$.
    \textbf{(b)} Optimized ZYZ pulse parameters $(\alpha^\ast,\beta^\ast,\gamma^\ast)$ as functions of $\theta$ for the $D=1$ circuit at $L=20$.
    Experimentally, these tunable pulse angles can be used to construct $U_{\rm approx}$ and extract work from the subsystem.
    \textbf{(c)} Representative variational landscape at $\theta=0$ ($D=1$, $L=20$).
    The axes scan the physical pulse angles $\alpha_1/\pi$ and $\beta_1/\pi$, while $\gamma_1$ is fixed at its optimized value. The red star marks the pulse setting that maximizes $W_{\rm approx}$.
    The color encodes $W_{\rm approx}$, and the width of the bright region visualizes its sensitivity to pulse calibration. Here a broad high work basin indicates a robust extraction protocol, whereas a narrow bright peak would demand much more precise angle control.}
    \label{fig:compiled_theta_landscape}
\end{figure*}

Together, \cref{fig:compiled_circuit_depth,fig:compiled_theta_landscape} show that the ergotropy bound has a shallow-circuit counterpart: the extracted work improves systematically with depth and remains robust over a broad region of pulse parameters.

\subsection{Ergotropy scaling of finite energy shell in the inversion anti-symmetric and $\pi$ momentum sector}

\begin{figure*}[htbp]
    \centering
    \begin{overpic}[width=0.33\columnwidth]{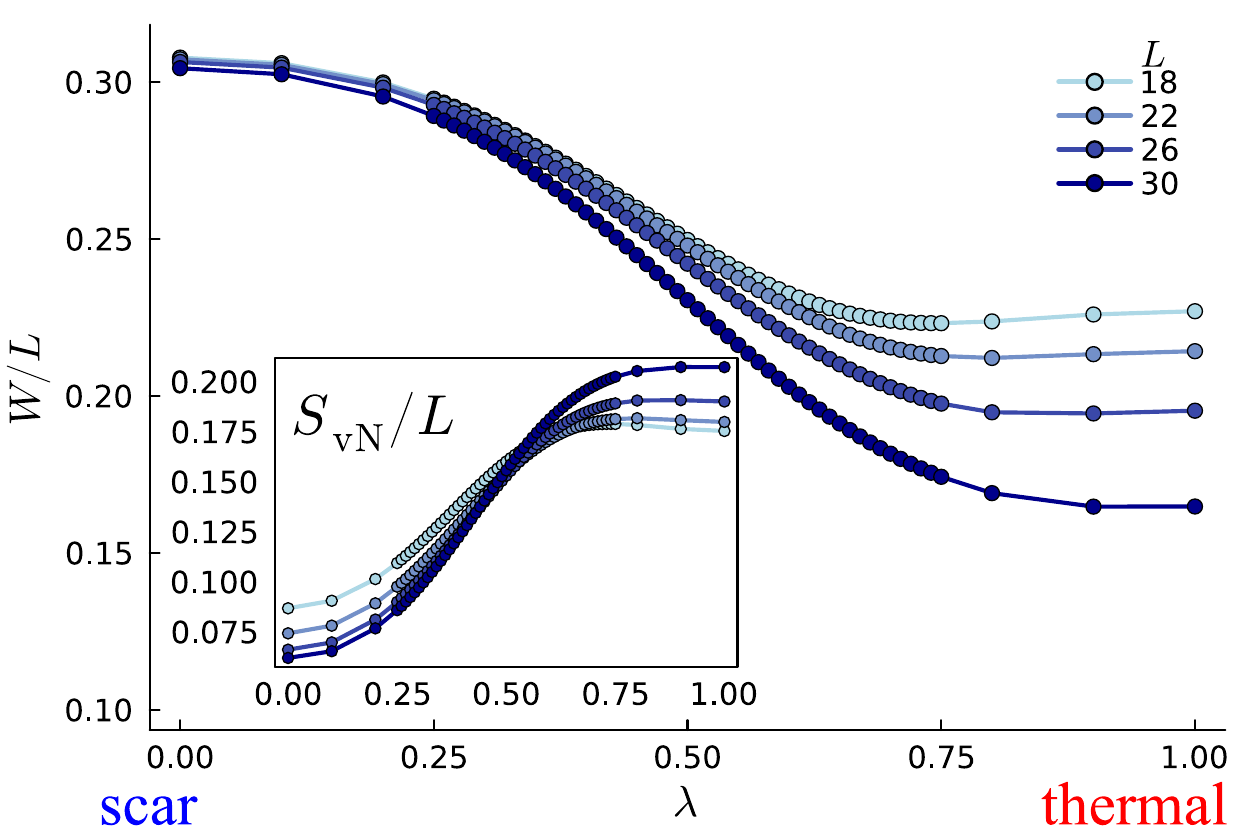}
        \put(-1,60){\textbf{(a)}}
    \end{overpic}
    \begin{overpic}[width=0.33\columnwidth]{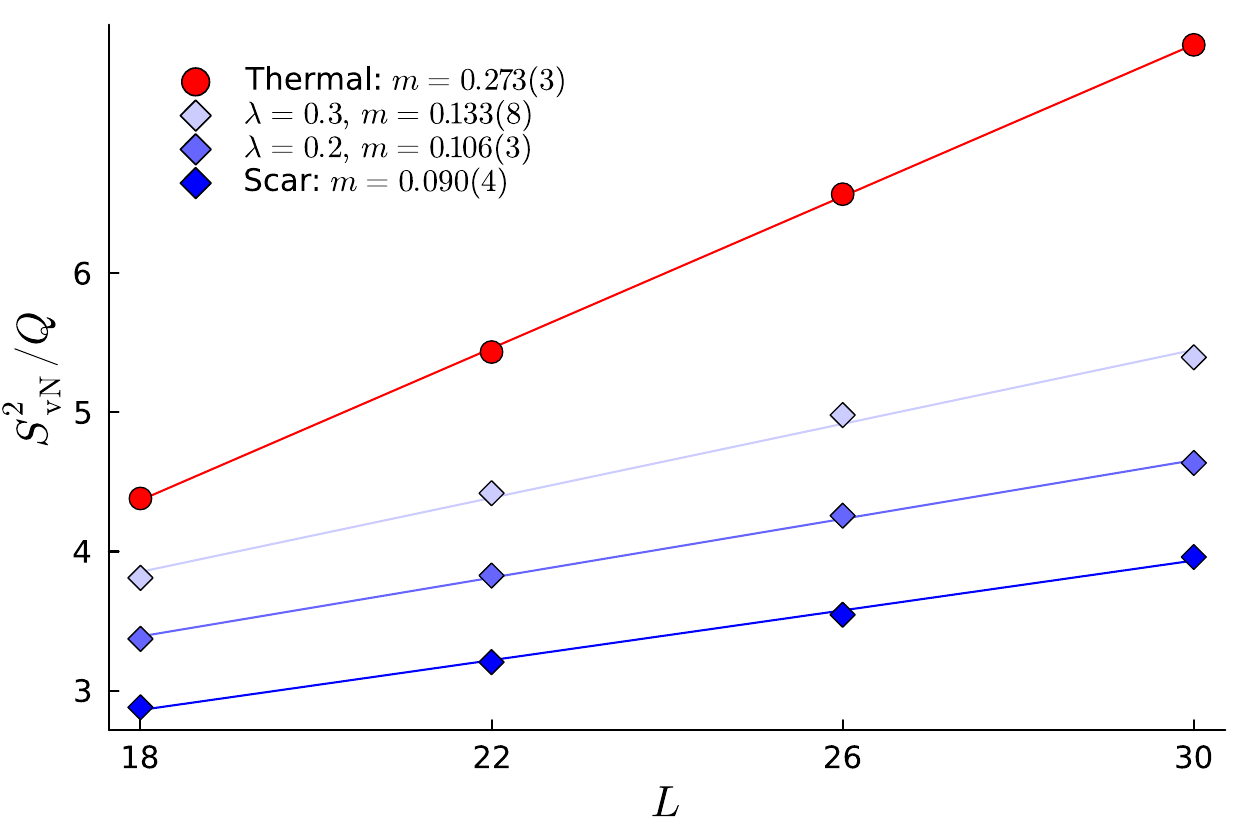}
        \put(-1,60){\textbf{(b)}}
    \end{overpic}
    \begin{overpic}[width=0.33\columnwidth]{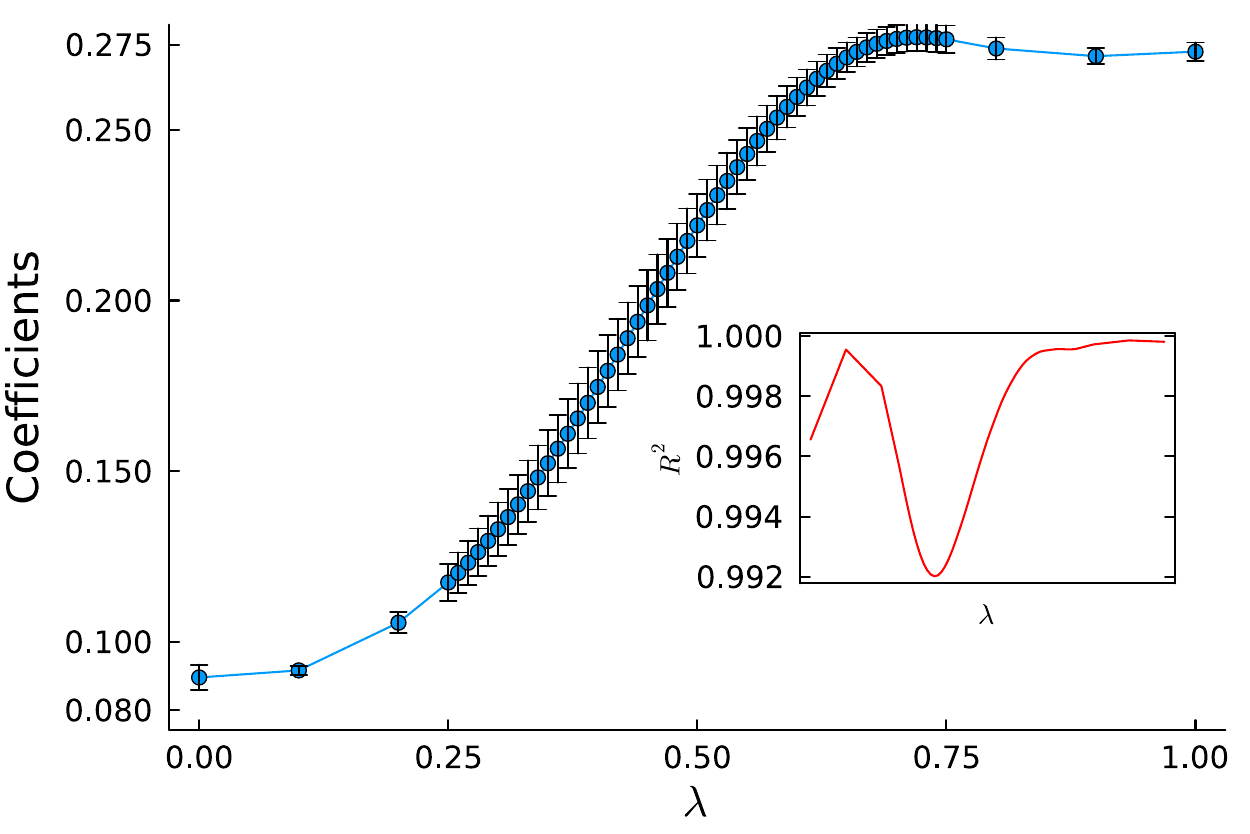}
        \put(-2,60){\textbf{(c)}}
    \end{overpic}
    \caption{{\bf Finite energy shell scar-thermal crossover in given symmetry sector}. 
    {\bf (a)} The ergotropy density $W/L$ and entanglement entropy density $S_{{\rm vN}}/L$ are averaged over the ensemble at finite energy shell $[E-\Delta E, E+\Delta E]$ in the inversion symmetric and zero momentum sector ($k = 0, I = 1$). $ E \approx 1.34 $ is set to be the shell of first positive energy scar state, which only lies in $(k, I) = (0, 1)$ sector for $L=4i+2$ system size. And $\Delta E \sim 1{\rm e}^{-3}$ for $N = 30$. The scaling behavior is extensive scaling for scar states and sub-extensive scaling for thermal states.
    {\bf (b)} The phenomenological relationship $S_{\mathrm{vN}}^2/Q = n + m L$ is fitted for system size $L=18$-$30$,
    {\bf (c)} The coefficient $m$ and fitting fidelity $R^2$ across different superposition $\lambda$ values for the finite energy shell, is consistent with the zero energy shell result shown in \cref{fig:scarthermalErgo}(b).
    }
    \label{fig:new_scar_scaling_symmetry}
\end{figure*}

The PXP model's algebraic structure features an approximate $\mathfrak{su}(2)$ algebra~\cite{EmergentSU2Scars} and, separately, an extensive zero-mode degeneracy originating from particle-hole and inversion symmetries~\cite{Number_zero_modes_PXP}. We want to clarify that the ergotropy scaling behavior is not a consequence of extensive zero modes, but rather stems from the dichotomy of entanglement entropy scaling between thermal and scar states, rooted in the emergent $\mathfrak{su}(2)$ algebraic structure. To avoid the extensive exact degeneracy of zero energy shell, we investigate the ergotropy scaling for the finite energy shell in the inversion symmetric and zero momentum symmetry sector for $L=4i+2$, where the first positive energy scar state appears. We choose a narrow finite energy shell $[E-\Delta E, E+\Delta E]$ around the first positive energy scar state and average the ergotropy $W$ and entanglement entropy $S_{\mathrm{vN}}$ of $\ket{{\rm scar}}$ and all $\ket{{\rm thermal}}$ states in this window. Although the energy among different eigenstates may lead to fluctuations in the ergotropy after lifting the degeneracy, such deviations can be suppressed and bounded by using finer energy windows for larger system sizes. As shown in \cref{fig:new_scar_scaling_symmetry}, the scaling behavior is consistent with the zero energy shell results shown in \cref{fig:scarthermalErgo}(a), with extensive scaling for scar states and sub-extensive scaling for thermal states, further corroborating the ergotropy-entanglement anti-correlation. We also derive similar phenomenological relationship $S_{\mathrm{vN}}^2/Q = n + m L$ for the finite energy shell, which is consistent with the zero energy shell result shown in \cref{fig:scarthermalErgo}(b). The fitting fidelity is listed in \cref{fig:new_scar_scaling_symmetry}(c). Thus our statement can be generalized to other models hosting scars, such as the AKLT model~\cite{AKLT_scar_analytical,AKLT_scar_numerical}, the Hubbard model~\cite{Eta_pairing_Yang1989}, and the spin-1 XY model~\cite{XY_scar_Iadecola2019}.

\clearpage
\section{Supplementary materials}

\subsection{Ergotropy and bound energy}

The thermodynamic characterization of quantum many-body states requires understanding how much stored energy can be extracted as work and how much remains inaccessible. For a many-body system where subsystem $A$ is described by the reduced density matrix $\rho_A$ and subsystem Hamiltonian $H_A$, the total subsystem energy $E = \mathrm{tr}(\rho_A H_A)$ (with subsystem ground state energy set to zero) decomposes into ergotropy $W$ and bound energy $Q$, satisfying $E = W + Q$. The ergotropy represents the maximum work extractable via unitary operations~\cite{ErgotropyOriginalAllahverdyan2004} without entropy production, while the bound energy quantifies the thermodynamically inaccessible portion that remains locked due to quantum correlations with the environment $\bar{A}$.

\begin{figure*}[htbp!]
    \centering
    \begin{overpic}[width=0.34\columnwidth]{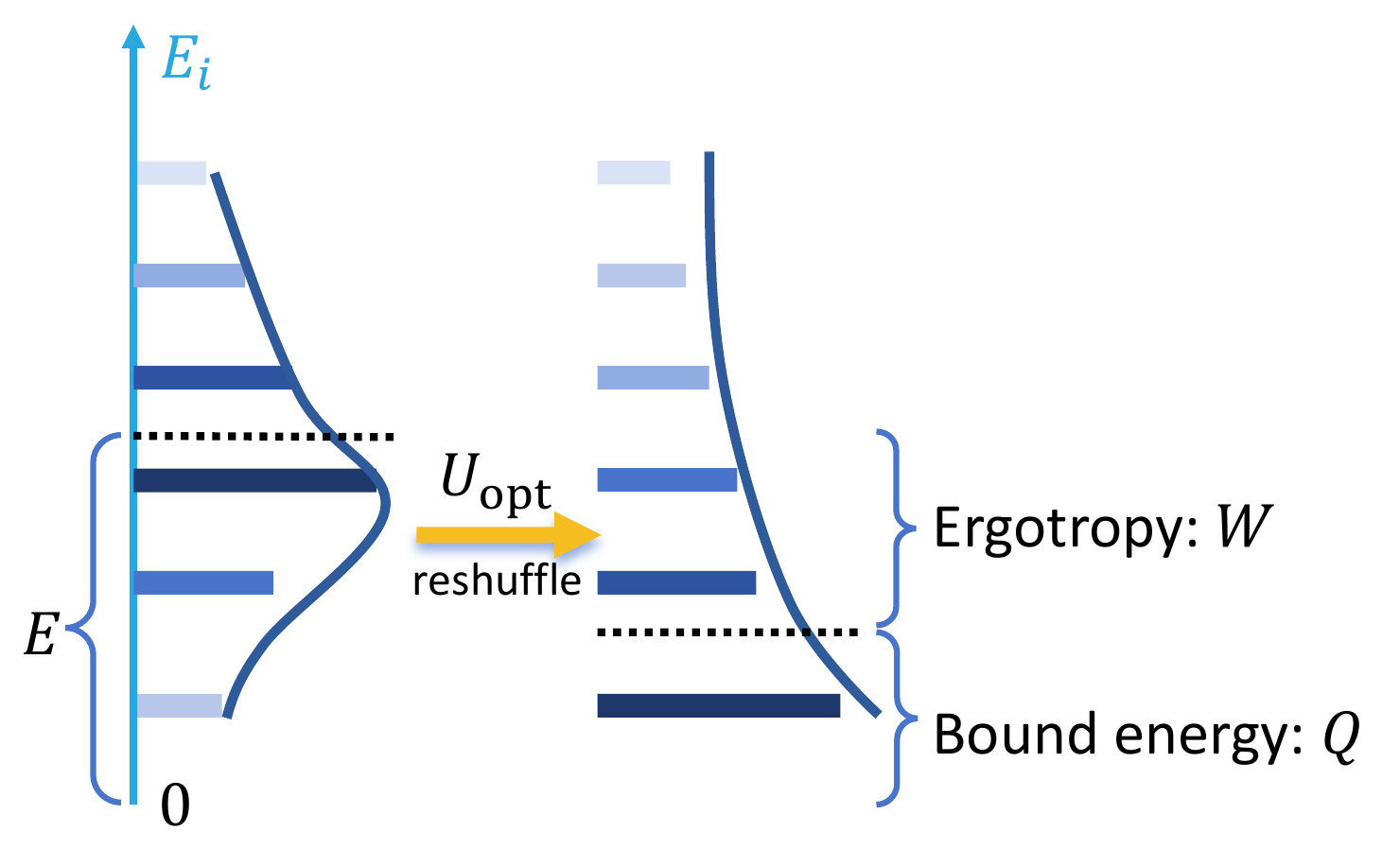}
        \put(-1,55){\textbf{(a)}}
    \end{overpic}
    \begin{overpic}[width=0.31\columnwidth]{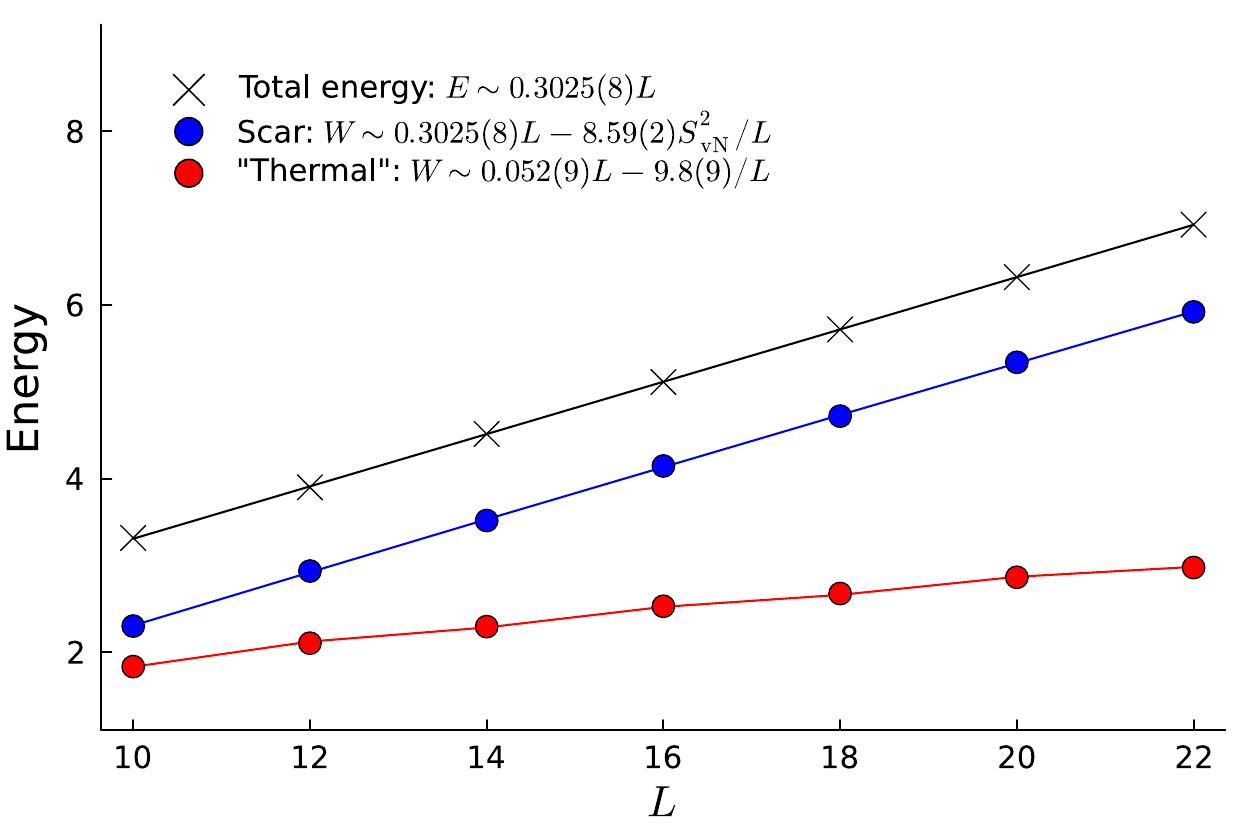}
        \put(-4,60){\textbf{(b)}}
    \end{overpic}
    \begin{overpic}[width=0.31\columnwidth]{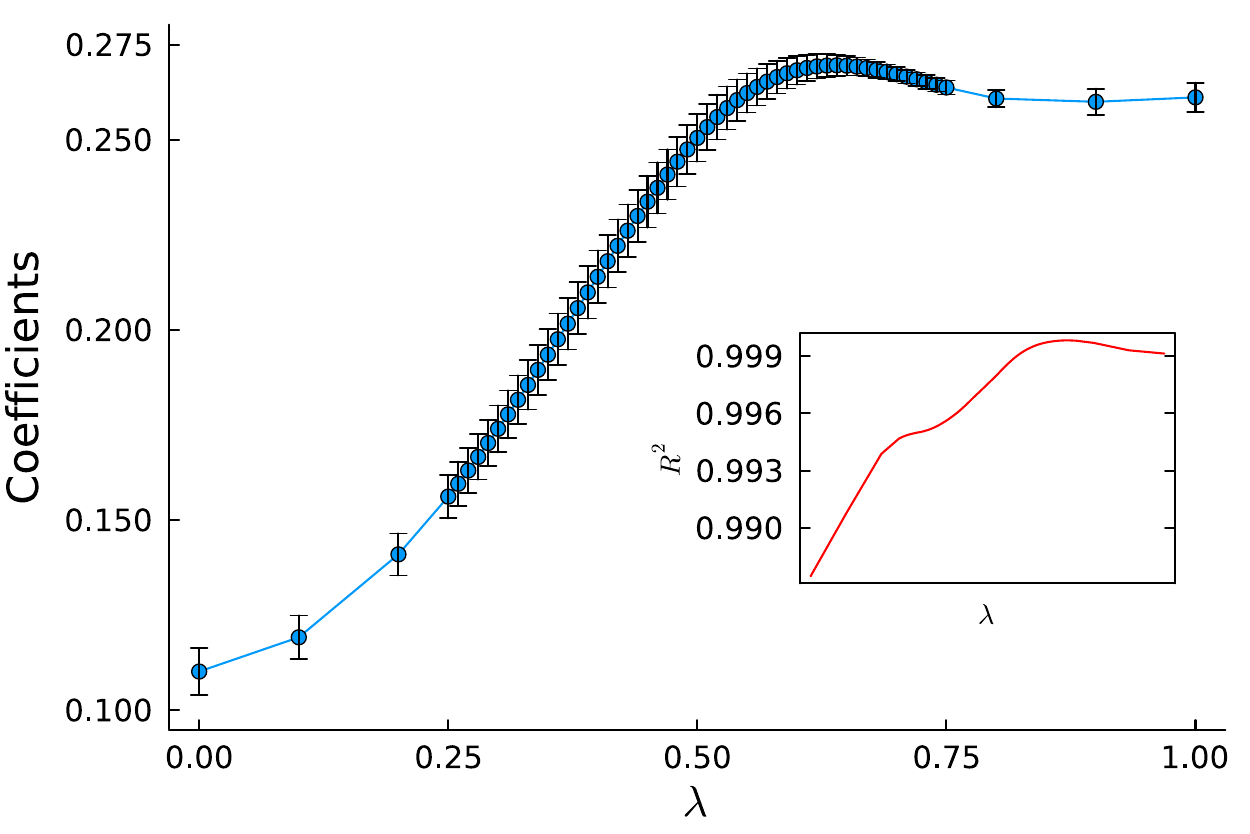}
        \put(-4,60){\textbf{(c)}}
    \end{overpic}
    \caption{
    \textbf{(a)} Schematic illustration of ergotropy definition and its relationship to bound energy. The density matrix eigenvalue population of subsystem state $\rho_A$ on energy spectrum $H_A$ (deep blue solid curve) is optimally redistributed by $U_{{\rm opt}}$ to anti-align with the subsystem energy spectrum (blue axis), maximizing work extraction $W$ while minimizing bound energy $Q$, with average energy drop (dash line). The gradient blues represents the magnitude of population. Here total energy satisfies $E = W + Q$.
    \textbf{(b)} System size scaling of ergotropy $W$ and total energy $E$ that is only supported on subsystem $A$: while total subsystem energy $E$ scales linearly with system size $L$, ergotropy exhibits extensive scaling $W \sim L-\ln^2(L)/L$ for scar states versus sub-extensive scaling $W \sim L^{-1}$ for thermal states across $L=10$--$22$.
    \textbf{(c)} The coefficient $m$ of the phenomenological relationship $S_{\mathrm{vN}}^2/Q = n + m L$, corresponding to the slope in \cref{fig:scarthermalErgo}(b), and fitting fidelity $R^2$ across different superposition $\lambda$ values.
    }
    \label{fig:ergotropy_vs_entropy}
\end{figure*}

The optimal work extraction protocol involves applying a unitary $U_{\rm opt}$ that reorders the population distribution of $\rho_A$ to anti-align with the energy spectrum of $H_A$, as illustrated in \cref{fig:ergotropy_vs_entropy}(a). This procedure transforms $\rho_A$ to its passive state, from which no further work can be extracted. The distinction between scar and thermal states manifests dramatically in their bound energy behavior: as shown in \cref{fig:scarthermalErgo}(b), the ratio $S^2_{\mathrm{vN}}/Q$ exhibits different system size dependence for the two cases. While the total subsystem energy $E$ scales extensively with system size $L$ in both cases, the ergotropy exhibits fundamentally different scaling laws. As shown in \cref{fig:ergotropy_vs_entropy}(b), scar states maintain extensive ergotropy $W \sim L - \ln^2(L)/L$, whereas thermal states yield sub-extensive scaling $W \sim L^{-1}$, approaching vanishing extractable work density in the thermodynamic limit. These two curves correspond to the $\lambda = 0$ and $\lambda = 1$ limits of ergotropy density in \cref{fig:scarthermalErgo}(a), from which we extrapolate to the thermodynamic limit. The residual ergotropy observed for thermal states reflects imperfect state separation and finite size effects.

\subsection{Many-body entanglement diagnostics}

Beyond bipartite entanglement entropy, multipartite entanglement measures provide complementary diagnostics for characterizing the crossover between scar and thermal regimes. We employ tripartite mutual information (TMI), bipartite mutual information (MI), and quantum Fisher information (QFI) density to probe different aspects of the entanglement structure across the interpolation parameter $\lambda$.

\begin{figure*}[htbp!]
    \centering
    \begin{overpic}[width=0.325\columnwidth]{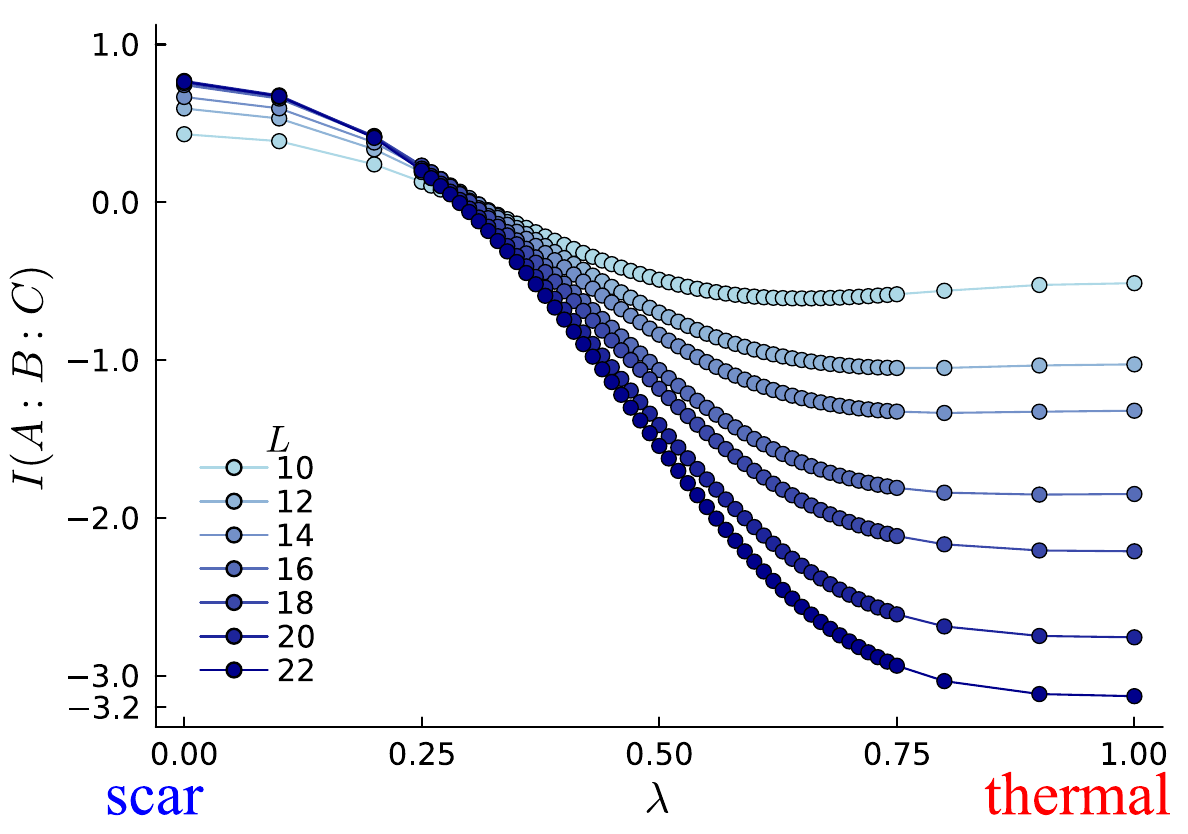}
        \put(-1,64){\textbf{(a)}}
    \end{overpic}
    \begin{overpic}[width=0.325\columnwidth]{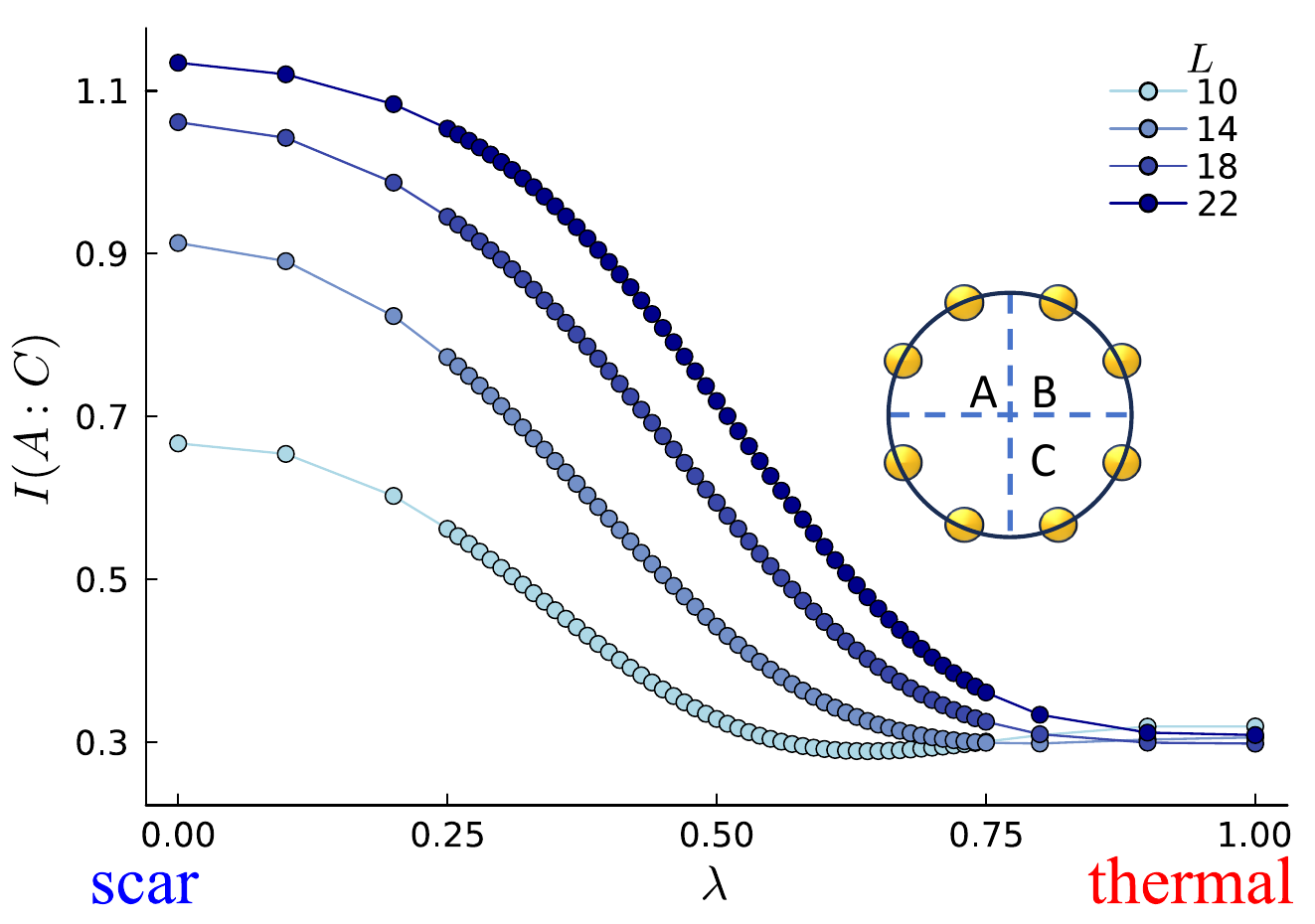}
        \put(-2,64){\textbf{(b)}}
    \end{overpic}
    \begin{overpic}[width=0.30\columnwidth]{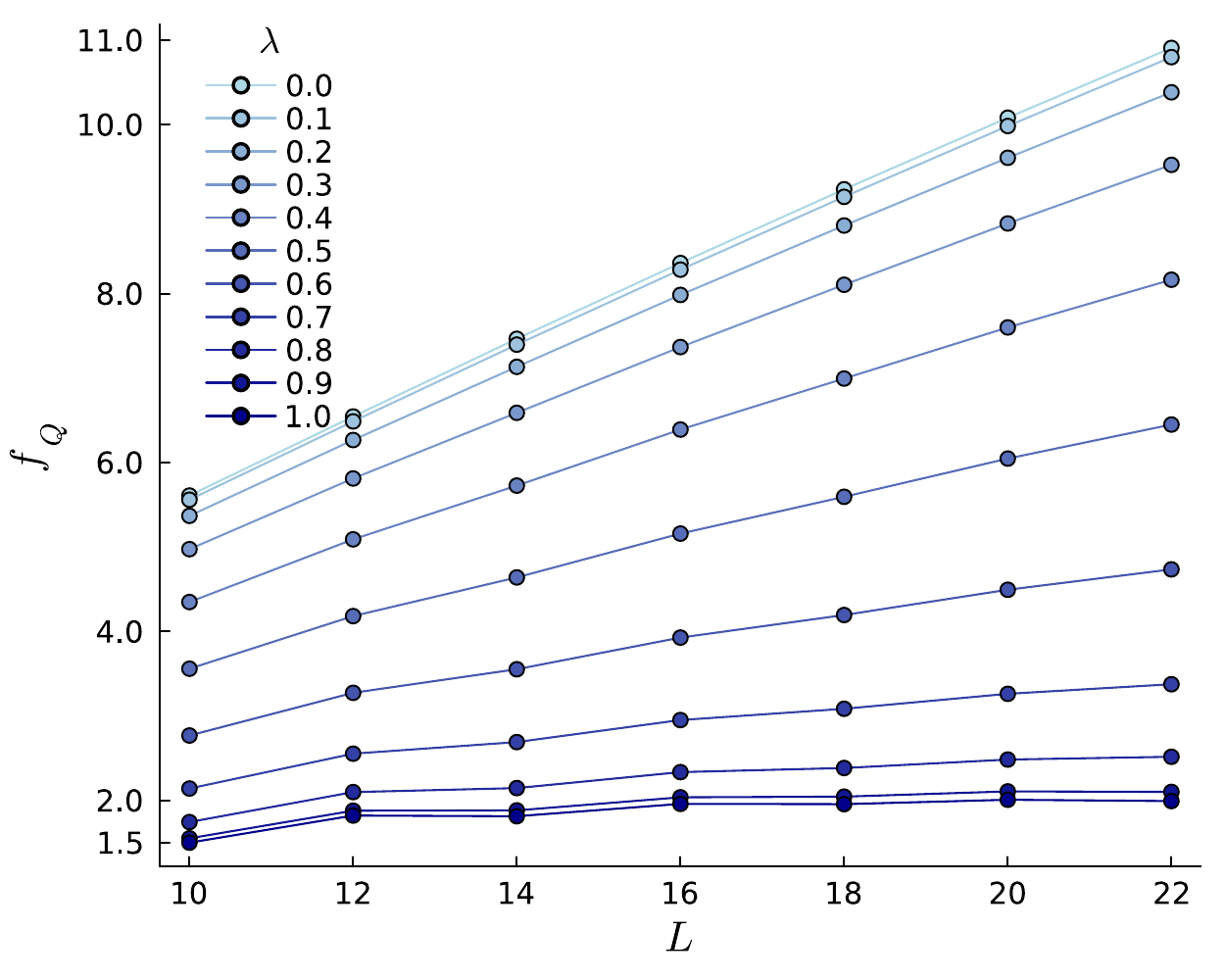}
        \put(-2,68){\textbf{(c)}}
    \end{overpic}
    \caption{\textbf{Many-body entanglement.}
    \textbf{(a)} Tripartite mutual information (TMI) $I(A:B:C)$ between intervals $A,B,C$ by partitioning system into four equal intervals as shown in (b), serves as a diagnostic for the entanglement crossover from area law to volume law phases. The transition exhibits scaling from constant to extensive behavior $I(A:B:C) \propto -L$, with pronounced even odd effects reflecting distinct entanglement structures for $L=4N$ versus $L=4N+2$ systems.
    \textbf{(b)} Bipartite mutual information (MI) $I(A:C)$ scaling shows the transition from extensive to sub-extensive behavior, with similar even odd effects, so we only show $L=4N+2$ data. 
    \textbf{(c)} System size dependence of QFI density across all superpositions: extensive scaling characterizes the scar regime ($\lambda=0$), while sub-extensive, approximately constant behavior characterizes the thermal regime ($\lambda=1$).
    }
    \label{fig:mutualinfofisherinfo}
\end{figure*}

The tripartite mutual information, defined as $I(A:B:C) = S_{{\rm vN}}(A) + S_{{\rm vN}}(B) + S_{{\rm vN}}(C) - S_{{\rm vN}}(AB) - S_{{\rm vN}}(BC) - S_{{\rm vN}}(AC) + S_{{\rm vN}}(ABC)$, quantifies genuine three-partite correlations and serves as a sensitive probe of topological entanglement entropy~\cite{Kitaev2006,Zeng2015QuantumIM}. For area law entangled states, TMI remains constant with system size, while volume law states exhibit extensive scaling $I(A:B:C) \propto -L$~\cite{Hosur_2016, Hayden_2013}. As shown in \cref{fig:mutualinfofisherinfo}(a), the crossover from scar ($\lambda = 0$) to thermal ($\lambda = 1$) regimes is clearly manifested in the TMI scaling behavior, with pronounced even odd effects arising from the distinct entanglement structures of $L = 4N$ versus $L = 4N+2$ systems under periodic boundary conditions.

The bipartite mutual information between non-adjacent intervals, $I(A:C) = S_{{\rm vN}}(A) + S_{{\rm vN}}(C) - S_{{\rm vN}}(AC)$, provides additional insight into long-range correlations. As shown in \cref{fig:mutualinfofisherinfo}(b), the MI transitions from extensive scaling in the scar regime to sub-extensive behavior in the thermal regime, consistent with the suppression of long-range correlations upon thermalization.

The quantum Fisher information density, $f_Q = \langle (O - \langle O \rangle)^2 \rangle / L$, evaluated with respect to the staggered magnetization operator $O = \sum_i (-1)^{i+1} Z_i$, witnesses genuine multipartite entanglement~\cite{ExtensiveQFI_pappalardi}. Remarkably, scar states exhibit extensive QFI density despite obeying area law entanglement entropy, indicating multipartite entanglement that is fundamentally distinct from the volume law entanglement of thermal states. The crossover in QFI density across system sizes $L = 10$ to $22$, shown in \cref{fig:mutualinfofisherinfo}(c), together with MI and TMI, collectively demonstrates that the scar-thermal transition extends beyond simple bipartite measures to encompass the full many-body entanglement structure, further corroborating the ergotropy-entanglement anti-correlation.

\subsection{Detailed relaxation dynamics}

The time resolved dynamics of ergotropy and entanglement provide microscopic insight into the mechanisms underlying the ergotropy entanglement anti-correlation observed in steady states. Here we present the full relaxation dynamics for representative quench protocols, comparing scarred ($\theta = 0$) and thermal ($\theta = \pi/2$) initial conditions across multiple system sizes.

\begin{figure*}[htbp!]
    \centering
    \begin{overpic}[width=0.36\columnwidth]{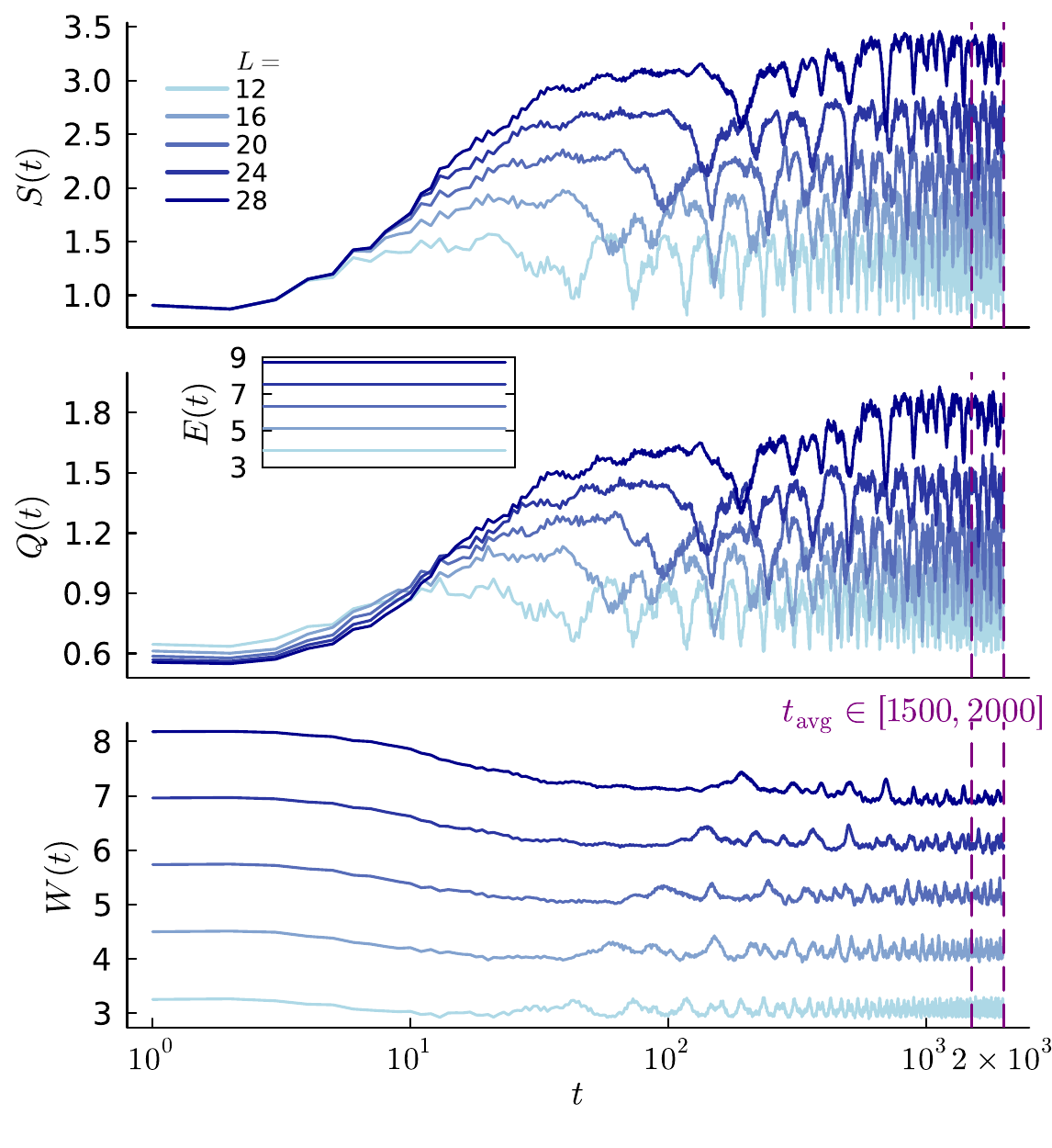}
        \put(-2,96){\textbf{(a)}}
    \end{overpic}
    \begin{overpic}[width=0.36\columnwidth]{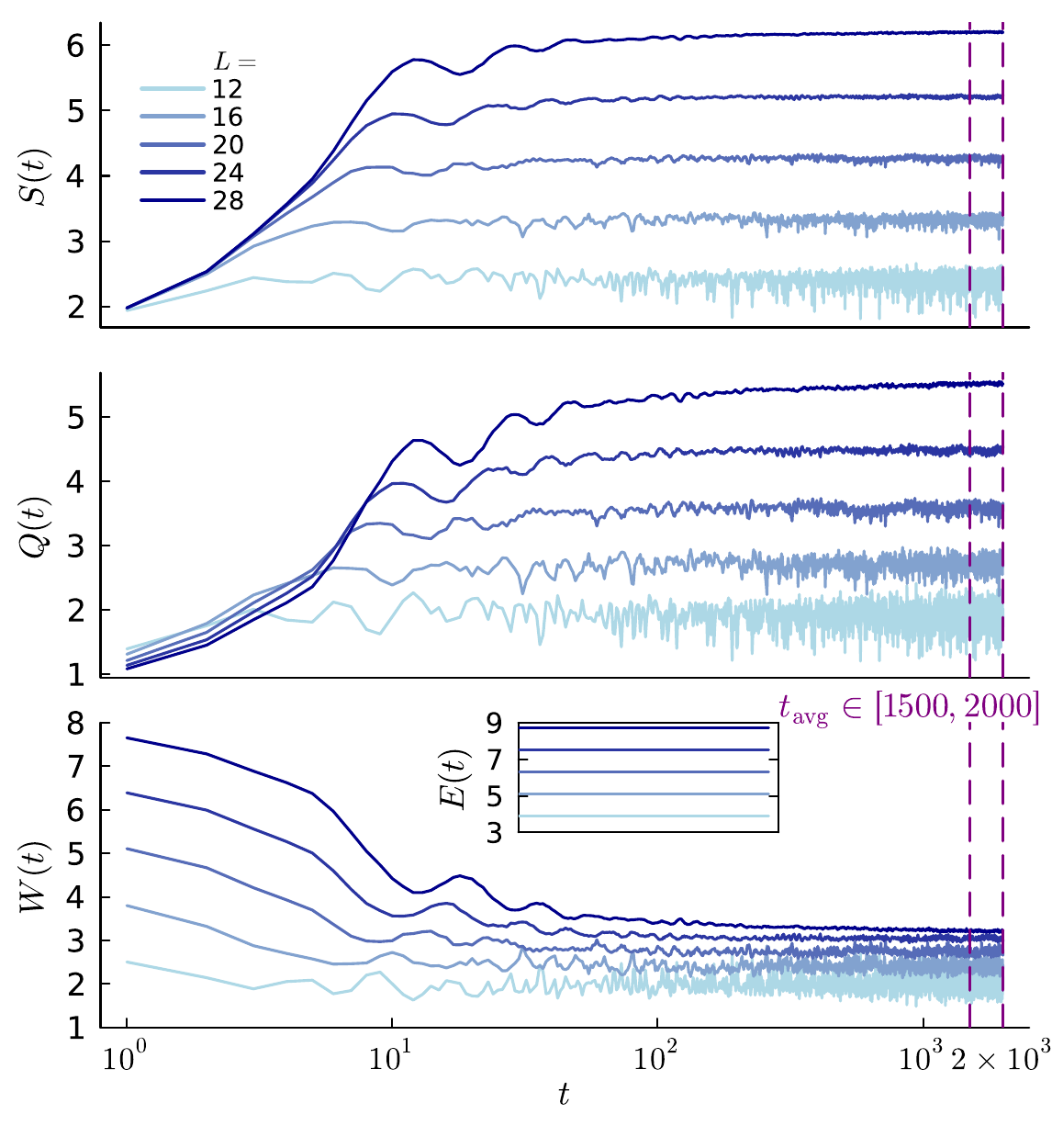}
        \put(-2,96){\textbf{(b)}}
    \end{overpic}
    \caption{
    \textbf{Quench dynamics evolution of ergotropy and entanglement entropy for different system sizes.}
    \textbf{(a)} Scarred quench dynamics ($\theta = 0$) reveal the microscopic origin of extensive work extraction capacity. The anti-correlated oscillations between $S_{\mathrm{vN}}(t)$ and $W(t)$, with synchronized behavior in bound energy $Q(t)$, demonstrate how low entanglement configurations enable high ergotropy states. The slow decay envelope preserves sizable extractable work even at late times, reflecting the non-thermal nature of scar relaxation.
    \textbf{(b)} Thermal quench dynamics ($\theta = \pi/2$) provide a stark contrast through rapid thermalization processes. The fast, monotonic decay of $W(t)$ occurs concomitantly with ballistic entanglement growth that saturates at volume law scaling. The bound energy $Q(t)$ rises quickly to approach its thermal maximum, reflecting the rapid approach to energetic passivity. This view of the relaxation dynamics complements the time averaged analysis presented in the main text and confirms that the entanglement ergotropy anti-correlation persists throughout the entire relaxation evolution across all system sizes studied.
    }
    \label{fig:additional_dynamics}
\end{figure*}

For the scarred quench shown in \cref{fig:additional_dynamics}(a), the dynamics exhibits several distinctive features. The ergotropy $W(t)$ and entanglement entropy $S_{\mathrm{vN}}(t)$ display persistent anti-correlated oscillations throughout the evolution, with peaks in $W(t)$ coinciding with dips in $S_{\mathrm{vN}}(t)$. This oscillatory behavior reflects the periodic revivals characteristic of scar dynamics, where the system coherently revisits low entanglement configurations that support high ergotropy within $\mathcal{H}_{{\rm scar}}$. The bound energy $Q(t)$ oscillates in phase with $S_{\mathrm{vN}}(t)$, consistent with the phenomenological relation $Q \propto S_{\mathrm{vN}}^2$ established in the main text. Crucially, the decay envelope of $W(t)$ is remarkably slow, preserving sizable extractable work even at late times $t \sim 10^3$. This non-thermal relaxation reflects the weak ergodicity breaking inherent to quantum many-body scars.

In stark contrast, the thermal quench dynamics shown in \cref{fig:additional_dynamics}(b) exhibits rapid thermalization. The ergotropy $W(t)$ undergoes fast decay concomitant with ballistic entanglement growth, with oscillations increasingly suppressed for larger system sizes. The entanglement entropy quickly saturates to volume law scaling $S_{\mathrm{vN}} \propto L$, while the bound energy $Q(t)$ rises toward its thermal maximum. The extractable work is efficiently suppressed within a short timescale $t \sim 10$, reflecting the rapid approach to energetic passivity characteristic of thermalizing systems. These contrasting dynamics confirm that the entanglement ergotropy anti-correlation persists throughout the entire relaxation process, providing a dynamical foundation for the steady state behavior presented in the main text.

\subsection{Separation of degenerate QMBS and thermal states}

In the middle of the PXP model spectrum, degenerate QMBS coexist with thermal states and undergo hybridization. Numerical diagonalization yields eigenstates $\ket{E_i}$ that are linear superpositions of pure scar and thermal components: $\ket{E_i} = \sum_j c_{ij} (\ket{\text{scar}} + \ket{\text{thermal}}_j)$. As the system size increases, this hybridization becomes more pronounced due to the exponentially growing density of thermal states, manifesting in anomalous physical quantities such as overlaps with the $\ket{\mathbb{Z}_2}$ state, entanglement entropy, and QFI~\cite{Papic18reviewScar,ExtensiveQFI_pappalardi}. To separate pure scar states from thermal states within degenerate subspaces, we employ the Forward Scattering Approximation (FSA) framework following Refs.~\cite{SGA_scar_Moudgalya2020,Papic21reviewscar,scars_structure}.

The separation procedure involves two key steps. First, we identify the degenerate subspace $\mathcal{H}_{\rm deg}$ where scar and thermal subspaces hybridize. We construct the projector onto this subspace as:
\begin{equation}
    P_{\mathrm{deg}} = \sum_{i=1}^{N_{\mathrm{deg}}} \ket{E_i}\bra{E_i} \,
\end{equation}
where $\ket{E_i}$ are degenerate eigenstates within the same energy shell $E$, and $N_{\mathrm{deg}}$ is the degeneracy. In our main text, we focus on the zero energy shell $P_{E=0}$.

Second, to describe the pure scar subspace $\mathcal{H}_{\rm scar} \subset \mathcal{H}$, we utilize spectrum generating algebra (SGA)~\cite{scars_structure,SGA_scar_Moudgalya2020} with raising operator $Q^{+}$ satisfying $Q^{+}\mathcal{H}_{\rm scar} \subset \mathcal{H}_{\rm scar}$ and:
\begin{equation}
    \left(\comm{H}{Q^{+}}  - \omega Q^{+}\right) \mathcal{H}_{\rm scar} = 0 \, ,
\end{equation}
where $\omega$ is the energy spacing between adjacent scar states. Starting from a base scar state $\ket{\mathcal{S}_0} \in \mathcal{H}_{\rm scar}$ with energy $E_0$, the complete scar tower is constructed as:
\begin{equation}
    \ket{\mathcal{S}_n} = (Q^{+})^n \ket{\mathcal{S}_0}, \quad E_n = E_0 + n \omega \, .
\end{equation}

In the PXP model, the SGA operators can be approximated using the FSA method, which constructs an approximate basis starting from the extremal weight state $\ket{\mathbb{Z}_2}$ with the generators:
\begin{equation}
    H^{\pm} = \sum_{i \in \text{even}} P_{i-1} \sigma_i^{\pm} P_{i+1} + \sum_{i \in \text{odd}} P_{i-1} \sigma_i^{\mp} P_{i+1} \ , \quad 
\end{equation}
and they satisfy $SU(2)$ algebra approximately:
\begin{equation}
    [H^{+}, H^{-}] \approx H^{z} \ , \quad
    [H^{z}, H^{\pm}] \approx \pm H^{\pm} \ ,
\end{equation}
where $H^{z} = \sum_{i \in \text{even}} P_{i-1} \sigma_i^{z} P_{i+1} - \sum_{i \in \text{odd}} P_{i-1} \sigma_i^{z} P_{i+1}$. The FSA basis states are constructed as:
\begin{equation}
    \ket{n} = \frac{(H^{+})^n \ket{\mathbb{Z}_2}}{||(H^{+})^n \ket{\mathbb{Z}_2}||} \ , \quad n=0,1,\cdots, L \ .
\end{equation}
Thus the projector for the FSA subspace $\mathcal{H}_{{\rm FSA}}$ can be written as:
\begin{equation}
    P_{\mathrm{FSA}} = \sum_{n=0}^{L} \ket{n}\bra{n} \ .
\end{equation}

\begin{figure*}[t!]
    \centering

    \begin{overpic}[width=0.32\columnwidth]{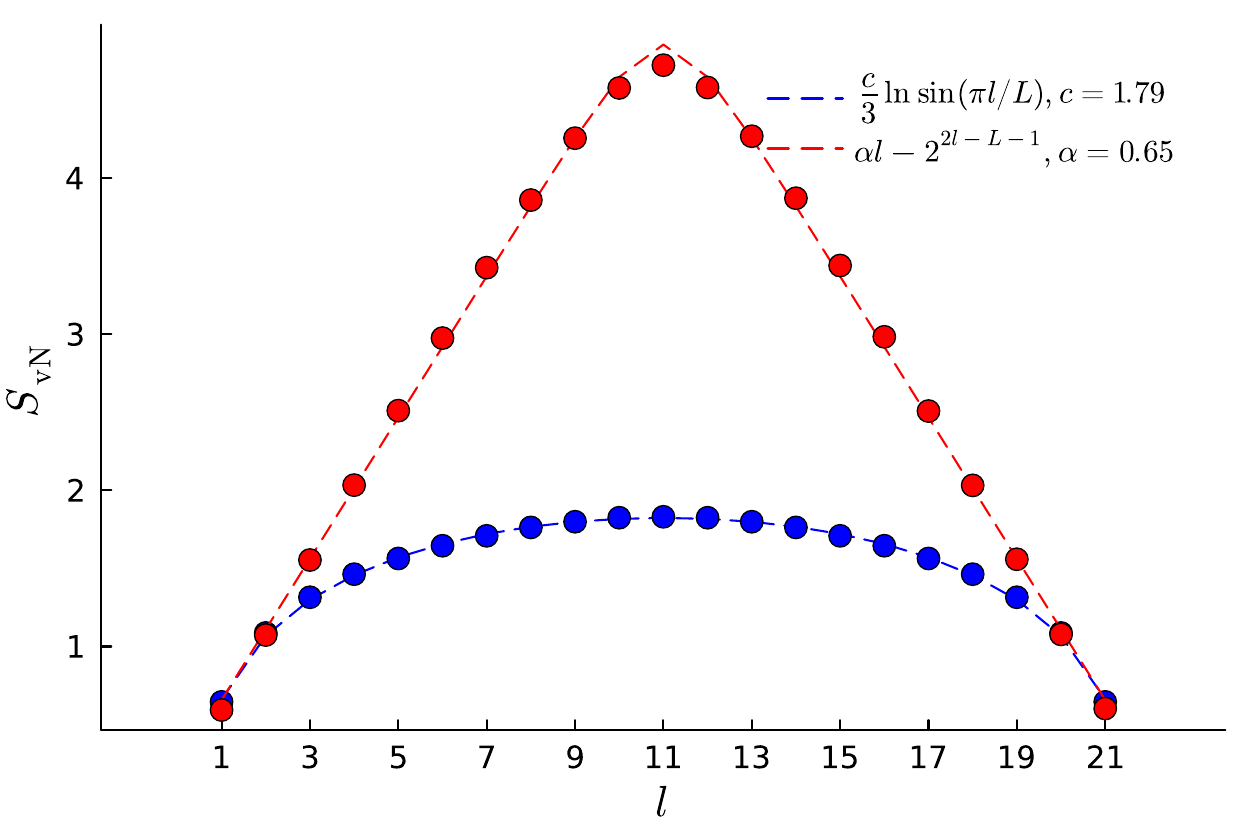}
        \put(-2,60){\textbf{(a)}}
    \end{overpic}
    \begin{overpic}[width=0.32\columnwidth]{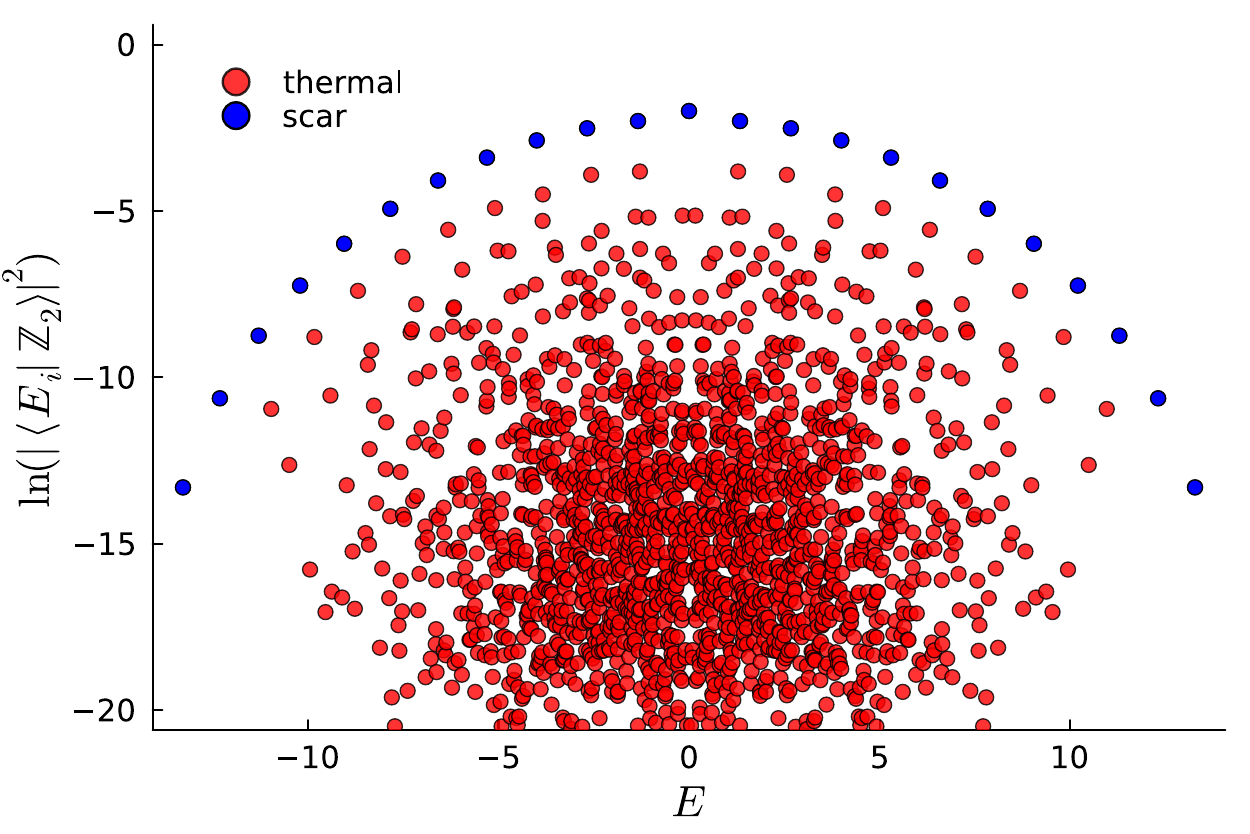}
        \put(-2,60){\textbf{(b)}}
    \end{overpic}
    \begin{overpic}[width=0.32\columnwidth]{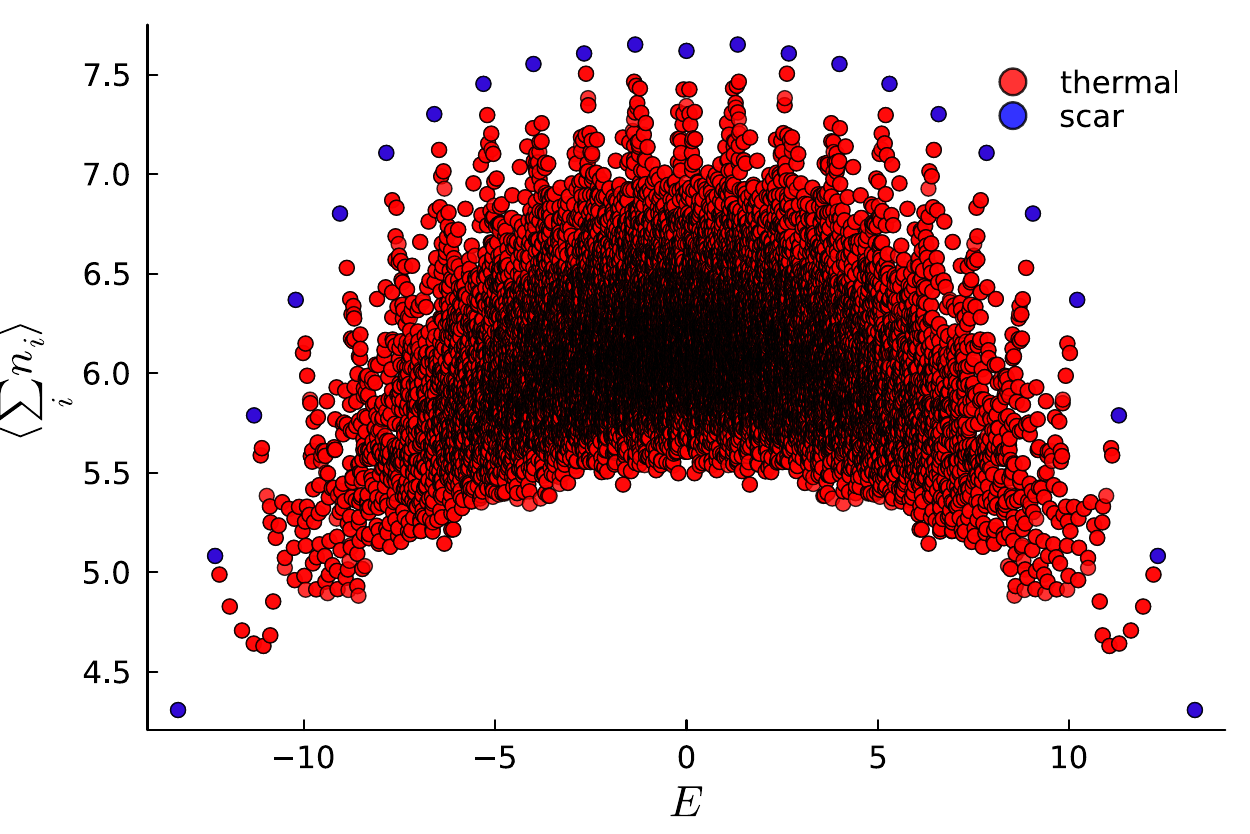}
        \put(-2,60){\textbf{(c)}}
    \end{overpic}
    \caption{{\bf Separated scar and thermal states} for the $L=22$ PXP model in the zero energy subspace $\mathcal{H}_{E=0}$, without restriction to specific symmetry sectors. 
    \textbf{(a)} Bipartite entanglement entropy $S_{\mathrm{vN}}$ as a function of subsystem size $l$ for the separated scar state and a representative thermal state in $\mathcal{H}_{E=0}$. The scar state (blue) exhibits sub-volume law scaling $S_{\mathrm{vN}} \sim \ln \sin(\pi l/L)$, in contrast to the Page curve of thermal states (red), $S_{\mathrm{vN}} \sim L - 2^{2l-L-1}$. Note that these two states are projected into the inversion symmetric sector.
    \textbf{(b)} Overlap $|\langle E | \mathbb{Z}_2 \rangle|^2$ between eigenstates and the $\ket{\mathbb{Z}_2}$ state across the energy spectrum. The large overlap of the zero energy scar is recovered after separation, while thermal states have negligible overlap, confirming the effectiveness of the method. Note that the sub-scar tower associated with $\ket{\mathbb{Z}_3}$ requires further separation.
    \textbf{(c)} Distribution of the local observable total particle numbers for the separated scar and thermal states. The scar state has a large anomalous values compared to the distribution predicted by the thermal ensemble, which exhibit a Gaussian distribution around the thermal average arc. The separation procedure effectively restores the anomalous properties of the scar state that were obscured by hybridization with thermal states in the degenerate subspace. Also note $E=0$ shell thermal states need to be projected into inversion symmetric sectors.
    }
    \label{fig:scar_thermal_separation}
\end{figure*}

We can separate the hybridization between $\mathcal{H}_{{\rm scar}}$ and $\mathcal{H}_{{\rm thermal}}$ by projecting the hybridized part onto $\mathcal{H}_{{\rm FSA}}$, because we expect that in $\mathcal{H}_{{\rm deg}}$ only one state is described by FSA, namely the scar. By diagonalizing the operator $P_{\mathrm{deg}} P_{\mathrm{FSA}} P_{\mathrm{deg}}$, 
\begin{equation}
    P_{\mathrm{deg}} P_{\mathrm{FSA}} P_{\mathrm{deg}} = \sum_j p_j \ket{p_j}\bra{p_j} \ ,
\end{equation}
the eigenstates $\ket{p_j}$ with the largest eigenvalue $p_j$ (approximately equal to 1) are expected to be scar states, while the rest with near zero eigenvalues correspond to thermal states. This method effectively separates scar states from thermal states within the degenerate subspace based on their overlap with the FSA subspace, as demonstrated in \cref{fig:scar_thermal_separation}, where the large overlap with $\ket{\mathbb{Z}_2}$ and low entanglement entropy are restored.

\subsection{Rotated states in constrained Hilbert space}
We construct constrained rotated states starting from $\ket{\mathbb{Z}_2} = \ket{1010\cdots}$ by applying local rotations $R(\theta) = e^{-i\theta Y/2}$ followed by projection $\mathbb{P}$ onto the constrained Hilbert space:
\begin{equation}
    \ket{\psi (\theta)} = \prod_i^L \tilde{R}_i \ket{\mathbb{Z}_2} = \mathbb{P} \ket{\phi (\theta)} = \prod_i^L \mathbb{P} R_i \ket{\mathbb{Z}_2} \ ,
\end{equation} 
where $\tilde{R} = \mathbb{P} R \mathbb{P}$ is the rotation operator in the constrained space.

Here we derive the explicit expression for these states in the constrained space and the symmetric basis. In the full Hilbert space, $R$ maps 
$\ket{0} \rightarrow \cos\frac{\theta}{2}\ket{0} + \sin\frac{\theta}{2}\ket{1}$ and
$\ket{1} \rightarrow -\sin\frac{\theta}{2}\ket{0} + \cos\frac{\theta}{2}\ket{1}$. Introducing $\gamma = \tan\frac{\theta}{2}$ for convenience, the rotated state is a product state: 
\begin{equation}
    \ket{\phi (\theta)}=\bigotimes_{i=1}^{N/2} \begin{pmatrix} 1 \\ \gamma \end{pmatrix}_i \bigotimes \begin{pmatrix} -\gamma \\ 1 \end{pmatrix}_i \cos^N\frac{\theta}{2} \ .
\end{equation} 
In the constrained subspace, the rotation becomes $\tilde{R} = \mathbb{P} R \mathbb{P}$, where the projection operator is $\mathbb{P} = \sum_i \ket{\tilde{\sigma}}_i \bra{\tilde{\sigma}}_i$ and $\ket{\tilde{\sigma}}$ denotes the basis states of the constrained space. Although writing down the matrix elements of $\tilde{R}$ explicitly is difficult, we can directly project the rotated state onto the subspace:
\begin{equation}
    \mathbb{P} \ket{\phi (\theta)}=\sum_i \ket{\tilde{\sigma}} \braket{\tilde{\sigma}}{\phi (\theta)} = \sum_i c_{\tilde{\sigma}} \ket{\tilde{\sigma}} \ .
\end{equation}

For $\braket{\tilde{\sigma}}{\phi (\theta)}$ in constraint basis, explicitly we have
\begin{equation}
\label{eq:projected_state}
c_{\tilde{\sigma}, \text{con}}= \gamma^{\text{\# 0@even}+\text{\# 1@odd}} (-1)^{\text{\# 0@even}} \cos^N\frac{\theta}{2} \ .
\end{equation}
At configuration $\ket{\tilde{\sigma}}$, the number of `0's at even sites and the number of `1's at odd sites in its binary string determine the amplitude.

If we continue one step forward to the symmetric constrained space, $\ket{\mathbb{Z}_2}_{\text{sym}}=\frac{1}{\sqrt{2}}(\ket{\bar{\mathbb{Z}}_2}+\ket{\mathbb{Z}_2})$. $\ket{\bar{\mathbb{Z}}_2} = \ket{0101\cdots}$ is the $\ket{\mathbb{Z}_2}$ translated by one site. Then $\ket{\phi (\theta)}_{\text{sym}} = R\ket{\mathbb{Z}_2}_{\text{sym}}=\frac{1}{\sqrt{2}}(\ket{\phi (\theta)}+T\ket{\phi (\theta)})$, $\mathbb{P}_{\text{sym}} \ket{\phi (\theta)}_{\text{sym}}=\sum_i \ket{\tilde{\sigma}_{\text{sym}}} \braket{\tilde{\sigma}_{\text{sym}}}{\phi (\theta)}_{\text{sym}}$. Here $\ket{\tilde{\sigma}_{\text{sym}}}$ is maximally symmetric constrained space's basis:
\begin{equation}
    \ket{\tilde{\sigma}_{\text{sym}}}=\frac{1}{\mathcal{N}}(\ket{\tilde{\sigma}_{k}}+I\ket{\tilde{\sigma}_{k}})=\frac{1}{\mathcal{N}}(\sum_i^N T^i \ket{\tilde{\sigma}_{r}} + I \sum_i^N T^i \ket{\tilde{\sigma}_{r}}) \ ,
\end{equation} 
where $\ket{\tilde{\sigma}_{r}}$ is the representative state and $\mathcal{N}$ is the overall normalization coefficient.

It is noted that $T^2 \ket{\phi (\theta)}=\ket{\phi (\theta)}$, so $\ket{\phi (\theta)}$ has zero or $\pi$ momentum $k=0, \pi$. Thus the inversion operation equals translation for $\ket{\phi (\theta)}$, $I \ket{\phi (\theta)}= T \ket{\phi (\theta)}$. So we only need to consider $\braket{\tilde{\sigma}_r}{\phi (\theta)}$ and $\braket{\tilde{\sigma}_r}{T \phi (\theta)}$. Similarly, 
\begin{equation}
\label{eq:projected_state_inv}
c_{T\tilde{\sigma}, \text{con}}= \gamma^{\text{\# 0@odd}+\text{\# 1@even}} (-1)^{\text{\# 0@odd}} \cos^N\frac{\theta}{2} \ .
\end{equation}

Finally, we derive the coefficient of $\ket{\phi (\theta)}$ in symmetric basis:
\begin{align*}
    c_{\tilde{\sigma}, \text{sym}} = \braket{\tilde{\sigma}_{\text{sym}}}{\phi (\theta)}_{\text{sym}} = \frac{2N}{\mathcal{N}}\braket{\tilde{\sigma}_{r}}{\phi (\theta)}_{\text{sym}} 
     = \frac{\sqrt{2}N}{\mathcal{N}} \left( c_{\tilde{\sigma}, \text{con}} + c_{T\tilde{\sigma}, \text{con}} \right) \ ,
\end{align*}
where $c_{\tilde{\sigma}, \text{con}}$ and $c_{T\tilde{\sigma}, \text{con}}$ are given by \cref{eq:projected_state,eq:projected_state_inv}, respectively.

The rotational symmetries of the constrained rotation operator yield:
\begin{equation}
    \begin{aligned}
 \tilde{R}(2\pi-\theta )&=X \tilde{R}(\theta)X \\ 
 \tilde{R}(\pi-\theta) &=X \tilde{R}(\theta)Z \ .
\end{aligned}
\end{equation}
These relations restrict our analysis to the interval $\theta \in [0, \pi/2]$.

\subsection{Energy and entanglement entropy for rotated states}

We start from the tensor network representation \cref{eq:state_tensor} of $\ket{\psi(\theta)}$, where the local on-site rotation $R(\theta)$ is applied to each qubit (either $\ket{1}$ or $\ket{0}$), followed by local projectors $P_{{\rm CZ}} = \frac{1+{\rm CZ}}{2}$ that enforce the constraint of no two adjacent excited atoms. The Rydberg blockade projector $P_{{\rm CZ}}$ acts on bonds with the matrix form $P_{{\rm CZ}} = \begin{pmatrix} 1 & 1 \\ 1 & 0 \end{pmatrix}$. We incorporate the on-site rotations $R(\theta)\ket{0}$ and $R(\theta)\ket{1}$ 
\begin{equation*}
        \begin{pmatrix}
            -\sin\frac{\theta}{2} & 0 \\
            0 & \cos\frac{\theta}{2}
        \end{pmatrix}, \ \begin{pmatrix}
            \cos\frac{\theta}{2} & 0 \\
            0 & \sin\frac{\theta}{2}
        \end{pmatrix}
\end{equation*}
into the bond tensors as weight. Thus we denote the bond tensors as A, B:
\begin{equation}
    A= \begin{pmatrix}
        -\sin\frac{\theta}{2} & -\sin\frac{\theta}{2} \\
        \cos\frac{\theta}{2} & 0 
    \end{pmatrix}, \quad B=\begin{pmatrix}
        \cos\frac{\theta}{2} & \cos\frac{\theta}{2} \\
        \sin\frac{\theta}{2} & 0
    \end{pmatrix} \ .
\end{equation}

The transfer matrix building block $E_{AB}$ is given by:
\begin{equation}
     E_{AB}=\begin{pmatrix}
        & \sin^2\frac{\theta}{2} &\cos^2\frac{\theta}{2}\sin^2\frac{\theta}{2}  & \cos^2\frac{\theta}{2}\sin^2\frac{\theta}{2} & \cos^2\frac{\theta}{2}\sin^2\frac{\theta}{2}\\
        & 0 & 0 & 0 & 0 \\
        & 0 & 0 & 0 & 0 \\
        & \cos^4\frac{\theta}{2}  & \cos^4\frac{\theta}{2} & \cos^4\frac{\theta}{2} & \cos^4\frac{\theta}{2}
    \end{pmatrix}
\end{equation}
obtained by elementwise multiplication of $A$ and $B$, followed by concatenation of two Z spider isometries. Similarly, $E_{BA}$ is:
\begin{equation}
    E_{BA}=\begin{pmatrix}
        & \cos^2\frac{\theta}{2} & \cos^2\frac{\theta}{2}\sin^2\frac{\theta}{2} &\cos^2\frac{\theta}{2}\sin^2\frac{\theta}{2} & \cos^2\frac{\theta}{2}\sin^2\frac{\theta}{2} \\
        & 0 & 0 & 0 & 0 \\
        & 0 & 0 & 0 & 0 \\
        & \sin^4\frac{\theta}{2} & \sin^4\frac{\theta}{2} & \sin^4\frac{\theta}{2} & \sin^4\frac{\theta}{2}
    \end{pmatrix}
\end{equation}

If we denote $f(\theta) = \sqrt{2} \sqrt{44 \cos(2\theta) - 3 \cos(4\theta) + 87}$, then $E_{AB}$ and $E_{BA}$ share two common non-zero eigenvalues: 
\begin{equation}
    \lambda_1 = \frac{1}{32}\left(2\cos(2\theta)+f(\theta)+14\right), \
    \lambda_2 = \frac{1}{32}\left(2\cos(2\theta)-f(\theta)+14\right) \ , 
\end{equation}
which can be canonicalized to $1$ and $\frac{\lambda_2}{\lambda_1}$. As expected from intuition, their left and right eigenstates $\ket{\lambda_{i, AB}^L}, \ket{\lambda_{i, AB}^R}$ satisfy sublattice reflection symmetry:
\begin{equation}
    \ket{\lambda_{i, AB}^L(\theta)} = \ket{\lambda_{i, BA}^L(\pi - \theta)}, \quad \ket{\lambda_{i, AB}^R(\theta)} = \ket{\lambda_{i, BA}^R(\pi - \theta)} \ .
\end{equation}

As for the energy $E(\theta)=\langle \psi(\theta)|H|\psi(\theta)\rangle$, the MPO transfer matrix is constructed from transfer matrix $E^O$ after incorporating $PXP$ into the transfer matrix. Due to the constraint of projector $P_{{\rm CZ}}$, only $|00\rangle$ input is allowed, thus $E^O_{BA}$ and $E^O_{AB}$ are:
\begin{equation}
    E^O_{BA}=\begin{pmatrix}
        & -2\cos^3\frac{\theta}{2}\sin\frac{\theta}{2} & 0 & 0 & 0 \\
        & 0 & 0 & 0 & 0 \\
        & 0 & 0 & 0 & 0 \\
        & 0 & 0 & 0 & 0
    \end{pmatrix}, \ E^O_{AB}=\begin{pmatrix}
        & 2\sin^3\frac{\theta}{2}\cos\frac{\theta}{2} & 0 & 0 & 0 \\
        & 0 & 0 & 0 & 0 \\
        & 0 & 0 & 0 & 0 \\
        & 0 & 0 & 0 & 0
    \end{pmatrix}
\end{equation}

\begin{figure}[hbtp!]
    \centering
    \begin{overpic}[width=0.4\columnwidth]{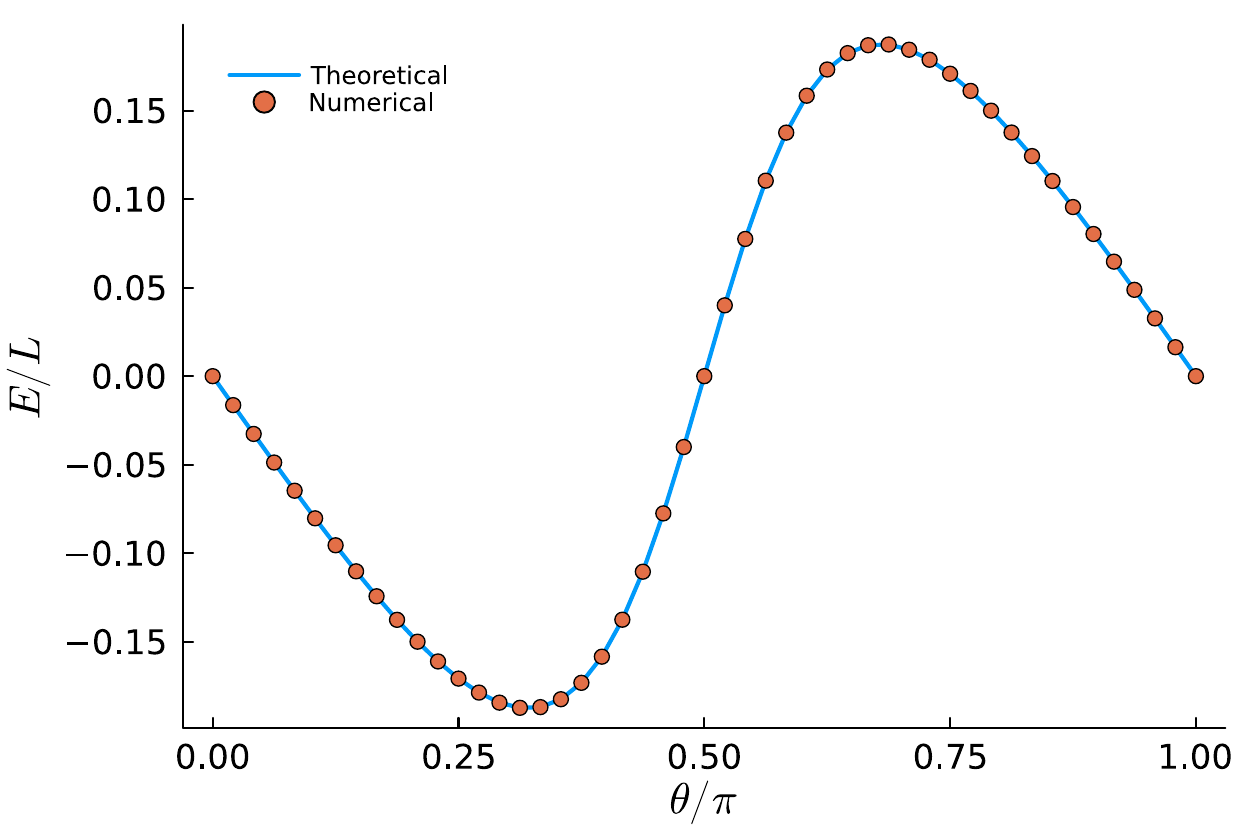}
        \put(-2,60){\textbf{(a)}}
    \end{overpic}
    \begin{overpic}[width=0.4\columnwidth]{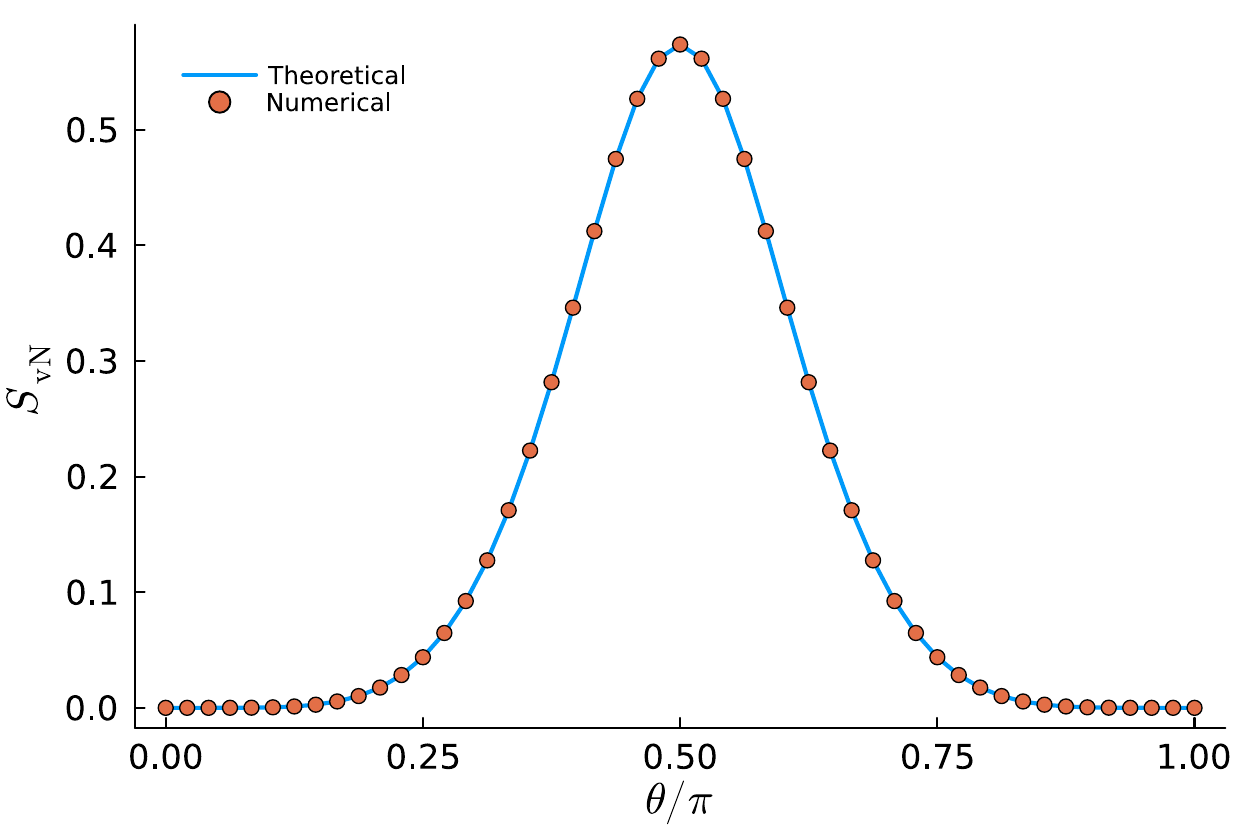}
        \put(-2,60){\textbf{(b)}}
    \end{overpic}
    \caption{{\bf Energy density and entanglement entropy of rotated initial states.}
    \textbf{(a)} Energy density $E/L$ of $\ket{\psi(\theta)}$ computed from \cref{eq:energy_density} (solid line) compared with numerical exact diagonalization (symbols), showing excellent agreement across all rotation angles $\theta$.
    \textbf{(b)} Entanglement entropy $S_{{\rm vN}}$ of $\ket{\psi(\theta)}$ for half-chain bipartition. Analytical calculations using the transfer matrix method (solid line) match numerical results (symbols) perfectly.
    } 
    \label{fig:state_energy}
\end{figure}

Note that the distinction between subsystem energy and half-chain energy under PBC becomes negligible in the thermodynamic limit. For simplicity, we consider systems with $L = 4n$ sites (integer $n$), yielding $2n$ copies each of $E_{BA}$ and $E_{AB}$ building blocks. The energy density is calculated as:
\begin{equation}\label{eq:energy_density}
    \begin{aligned}
        &E(\theta)/2L =\langle \phi(\theta)|H|\phi(\theta)\rangle /2L \\
        &= \frac{\sum_{i=1}^{2n}\tr(E_{BA}^{2n-1}E_{BA}^{O}) +\tr(E_{AB}^{2n-1}E_{AB}^{O})  }{\tr(E^{2n}) ~ 2L}  \\
        =\frac{1}{4(\lambda_1^{2n}+\lambda_2^{2n})}\sum_{i=1,2}& \lambda_i^{2n-1}\bra{\lambda_{i,BA}^L}E^O_{BA}\ket{\lambda_{i,BA}^R} +\frac{1}{4(\lambda_1^{2n}+\lambda_2^{2n})}\sum_{i=1,2}\lambda_i^{2n-1}\bra{\lambda_{i,AB}^L}E^O_{AB}\ket{\lambda_{i,AB}^R}  \\
    \end{aligned}   
\end{equation}
Computing \cref{eq:energy_density} yields the theoretical energy density shown in \cref{fig:state_energy}(a).

The entanglement spectrum of the reduced density matrix with a single entanglement cut can be constructed from the left and right dominant eigenstates $\ket{\lambda^R_{1, AB}}$ and $\ket{\lambda^L_{1, AB}}$~\cite{MPS_entanglement_spectrum}:
\begin{equation}
    {\rm eig}(\rho_A) = {\rm eig}(\lambda^R_{1, AB}\lambda^L_{1, AB}) \ ,
\end{equation}
where $\lambda^R_{1, AB}\lambda^L_{1, AB}$ is a $2\times 2$ matrix obtained by reshaping $\ket{\lambda^R_{1, AB}}\bra{\lambda^L_{1, AB}}$. For the case of two entanglement cuts, the entanglement spectrum is ${\rm eig}((\lambda^R_{1, AB}\lambda^L_{1, AB})^{\otimes 2})$, effectively doubling the entanglement entropy. The four eigenvalues are:
\begin{equation}
    \frac{1}{2}-h(\theta)+\sqrt{\frac{1}{4}-h(\theta)} \ , \frac{1}{2}-h(\theta)-\sqrt{\frac{1}{4}-h(\theta)} \ , h(\theta) \ , h(\theta) .
\end{equation}
If we denote $h(\theta) =  \frac{128\sin^6\theta}{(2\cos(2\theta) + f(\theta)+14)f^2(\theta)}$, then \cref{fig:state_energy}(b) shows the entanglement entropy calculated from this entanglement spectrum, matching perfectly with numerical calculations.

\subsection{Subsystem energy conservation for $L = 4N$ systems}

For systems with $L = 4N$ sites, the subsystem energy $E_A(\theta, t)$ is conserved during time evolution for all rotation angles $\theta$. This conservation law follows from analyzing the time derivative of the subsystem energy expectation value:

\begin{equation}\label{Equ:sub_energy_conservation}
    \begin{aligned}
        E_A(\theta, t) &= \langle \phi(\theta)| \mathbb{P} U^\dagger H_A  U \mathbb{P}| \phi(\theta) \rangle \\
        \dot{E}_A(\theta, t) &= \langle \phi(\theta)| \mathbb{P} [H_A, H] \mathbb{P}| \phi(\theta) \rangle  = \sum_{\tilde{\sigma}, \tilde{\rho}} c_{\tilde{\sigma}} c_{\tilde{\rho}} \langle \tilde{\sigma}| [H_A, H] |\tilde{\rho} \rangle = 0
    \end{aligned}
\end{equation}

where $H = H_A + H_B + H_{\text{int}}$ is the total Hamiltonian and coefficients $c_{\tilde{\sigma}}, c_{\tilde{\rho}}$ are given by \cref{eq:projected_state}. The commutator $[H_A, H]$ evaluates to:

\begin{equation}
    \begin{aligned}
[H_{A},H_{\text{int}}] & =[X_{1}P_{2},P_{1}X_{2}]P_{3}P_{4N}+[X_{1},P_{1}]P_{2}P_{4N-1}P_{4N}+ \\
 & +P_{2N-2}[X_{2N-1}P_{2N},P_{2N-1}X_{2N}]P_{2N+1} +P_{2N-1}[X_{2N},P_{2N}]X_{2N+1}P_{2N+2}
\end{aligned}
\end{equation}
For $\theta = 0$ or $\pi$, the projectors $P_{\text{even}}$ and $P_{\text{odd}}$ acting on $\ket{\mathbb{Z}_2}$ or $\ket{\bar{\mathbb{Z}}_2}$ yield zero directly. For generic $\theta \neq 0$, the $4N$ configurations contain opposing contributions from $\tilde{\sigma}$ and $\tilde{\rho}$ that cancel by symmetry, resulting in $\langle [H_A, H] \rangle = 0$.

\end{document}